\documentclass[twocolumn]{aastex631}

\usepackage{graphicx,textcomp,fancyhdr,hyperref,xcolor,fontawesome}
\definecolor{linkcolor}{rgb}{0.1216,0.4667,0.7059}
\usepackage[caption=false]{subfig}
\usepackage{float}
\usepackage{footnote}

\providecommand{\bjdtdb}{\ensuremath{\rm {BJD_{TDB}}}}

\providecommand{\msun}{\ensuremath{\,M_\Sun}}
\providecommand{\rsun}{\ensuremath{\,R_\Sun}}
\providecommand{\lsun}{\ensuremath{\,L_\Sun}}
\providecommand{\mj}{\ensuremath{\,M_{\rm J}}}
\providecommand{\rj}{\ensuremath{\,R_{\rm J}}}

\providecommand{\fave}{\langle F \rangle}

\begin{document}

\title{The TESS-Keck Survey. XIX. A Warm Transiting Sub-Saturn Mass Planet and a non-Transiting Saturn Mass Planet Orbiting a Solar Analog}

\author[0000-0002-0139-4756]{Michelle L. Hill}
\altaffiliation{NASA FINESST Fellow}
\affiliation{Department of Earth and Planetary Sciences, University of California, Riverside, CA 92521, USA}

\author[0000-0002-7084-0529]{Stephen R. Kane}
\affiliation{Department of Earth and Planetary Sciences, University of California, Riverside, CA 92521, USA}

\author[0000-0002-4297-5506]{Paul A. Dalba}
\altaffiliation{Heising-Simons 51 Pegasi b Postdoctoral Fellow}
\affiliation{Department of Astronomy and Astrophysics, University of California, Santa Cruz, CA 95064, USA}

\author[0000-0003-2562-9043]{Mason MacDougall}
\affiliation{Department of Physics \& Astronomy, University of California Los Angeles, Los Angeles, CA 90095, USA}

\author[0000-0002-3551-279X]{Tara Fetherolf}
\affiliation{Department of Physics, California State University, San Marcos, CA 92096, USA}
\affiliation{Department of Earth and Planetary Sciences, University of California, Riverside, CA 92521, USA}

\author[0000-0002-4860-7667]{Zhexing Li}
\affiliation{Department of Earth and Planetary Sciences, University of California, Riverside, CA 92521, USA}

\author[0000-0001-9771-7953]{Daria Pidhorodetska}
\altaffiliation{NASA FINESST Fellow}
\affiliation{Department of Earth and Planetary Sciences, University of California, Riverside, CA 92521, USA}

\author[0000-0002-7030-9519]{Natalie M. Batalha}
\affiliation{Department of Astronomy and Astrophysics, University of California, Santa Cruz, CA 95060, USA}

\author{Ian J. M. Crossfield}
\affiliation{Department of Physics \& Astronomy, University of Kansas, 1082 Malott, 1251 Wescoe Hall Dr., Lawrence, KS 66045, USA}

\author[0000-0001-8189-0233]{Courtney Dressing}
\affiliation{501 Campbell Hall, University of California at Berkeley, Berkeley, CA 94720, USA}

\author[0000-0003-3504-5316]{Benjamin Fulton}
\affiliation{NASA Exoplanet Science Institute/Caltech-IPAC, MC 314-6, 1200 E. California Blvd., Pasadena, CA 91125, USA}

\author[0000-0001-8638-0320]{Andrew W. Howard}
\affiliation{Department of Astronomy, California Institute of Technology, Pasadena, CA 91125, USA}

\author[0000-0001-8832-4488]{Daniel Huber}
\affiliation{Institute for Astronomy, University of Hawai`i, 2680 Woodlawn Drive, Honolulu, HI 96822, USA}
\affiliation{Sydney Institute for Astronomy (SIfA), School of Physics, University of Sydney, NSW 2006, Australia}

\author[0000-0002-0531-1073]{Howard Isaacson}
\affiliation{{Department of Astronomy,  University of California Berkeley, Berkeley CA 94720, USA}}
\affiliation{Centre for Astrophysics, University of Southern Queensland, Toowoomba, QLD, Australia}

\author[0000-0003-0967-2893]{Erik A Petigura}
\affiliation{Department of Physics \& Astronomy, University of California Los Angeles, Los Angeles, CA 90095, USA}

\author[0000-0003-0149-9678]{Paul Robertson}
\affiliation{Department of Physics \& Astronomy, University of California Irvine, Irvine, CA 92697, USA}

\author[0000-0002-3725-3058]{Lauren M. Weiss}
\affiliation{Institute for Astronomy, University of Hawai`i, 2680 Woodlawn Drive, Honolulu, HI 96822, USA}


\author[0000-0003-0012-9093]{Aida Behmard}
\altaffiliation{NSF Graduate Research Fellow}
\affiliation{Department of Astrophysics, American Museum of Natural History, 200 Central Park West, Manhattan, NY 10024, USA}

\author[0000-0001-7708-2364]{Corey Beard}
\altaffiliation{NASA FINESST Fellow}
\affiliation{Department of Physics \& Astronomy, University of California Irvine, Irvine, CA 92697, USA}

\author[0000-0003-1125-2564]{Ashley Chontos}
\altaffiliation{Henry Norris Russell Postdoctoral Fellow}
\affiliation{Department of Astrophysical Sciences, Princeton University, 4 Ivy Lane, Princeton, NJ 08540, USA}

\author[0000-0002-8958-0683]{Fei Dai} 
\altaffiliation{NASA Sagan Fellow}
\affiliation{Division of Geological and Planetary Sciences,
1200 E California Blvd, Pasadena, CA, 91125, USA}
\affiliation{Department of Astronomy, California Institute of Technology, Pasadena, CA 91125, USA}

\author[0000-0002-8965-3969]{Steven Giacalone}
\altaffiliation{NSF Astronomy and Astrophysics Postdoctoral Fellow}
\affiliation{Department of Astronomy, California Institute of Technology, Pasadena, CA 91125, USA}

\author[0000-0001-8058-7443]{Lea A.\ Hirsch}
\affiliation{Kavli Institute for Particle Astrophysics and Cosmology, Stanford University, Stanford, CA 94305, USA}

\author[0000-0002-5034-9476]{Rae Holcomb}
\affiliation{Department of Physics \& Astronomy, University of California Irvine, Irvine, CA 92697, USA}

\author[0000-0001-8342-7736]{Jack Lubin}
\affiliation{Department of Physics \& Astronomy, University of California Irvine, Irvine, CA 92697, USA}
\affiliation{Department of Physics \& Astronomy, University of California Los Angeles, Los Angeles, CA 90095, USA}

\author[0000-0002-7216-2135]{Andrew W. Mayo}
\affil{Department of Astronomy, University of California Berkeley, Berkeley, CA 94720, USA}

\author[0000-0003-4603-556X]{Teo Mo\v{c}nik}
\affiliation{Gemini Observatory/NSF's NOIRLab, 670 N. A'ohoku Place, Hilo, HI 96720, USA}

\author[0000-0001-8898-8284]{Joseph M. Akana Murphy}
\altaffiliation{NSF Graduate Research Fellow}
\affiliation{Department of Astronomy and Astrophysics, University of California, Santa Cruz, CA 95064, USA}

\author[0000-0001-7047-8681]{Alex S. Polanski}
\affil{Department of Physics and Astronomy, University of Kansas, Lawrence, KS 66045, USA}

\author{Lee J.\ Rosenthal}
\affiliation{Department of Astronomy, California Institute of Technology, Pasadena, CA 91125, USA}

\author[0000-0003-3856-3143]{Ryan A. Rubenzahl}
\altaffiliation{NSF Graduate Research Fellow}
\affiliation{Department of Astronomy, California Institute of Technology, Pasadena, CA 91125, USA}

\author[0000-0003-3623-7280]{Nicholas Scarsdale}
\affiliation{Department of Astronomy and Astrophysics, University of California, Santa Cruz, CA 95060, USA}

\author[0000-0002-1845-2617]{Emma V. Turtelboom}
\affiliation{Department of Astronomy, 501 Campbell Hall, University of California, Berkeley, CA 94720, USA}

\author[0000-0002-4290-6826]{Judah Van Zandt}
\affiliation{Department of Physics \& Astronomy, University of California Los Angeles, Los Angeles, CA 90095, USA}

\author[0000-0001-6637-5401]{Allyson~Bieryla} 
\affiliation{Center for Astrophysics ${\rm \mid}$ Harvard {\rm \&} Smithsonian, 60 Garden Street, Cambridge, MA 02138, USA}

\author[0000-0002-5741-3047]{David~ R.~Ciardi}
\affiliation{Caltech/IPAC-NASA Exoplanet Science Institute, 770 S. Wilson Avenue, Pasadena, CA 91106, USA}

\author[0000-0003-3773-5142]{Jason~D.~Eastman}
\affiliation{Center for Astrophysics ${\rm \mid}$ Harvard {\rm \&} Smithsonian, 60 Garden Street, Cambridge, MA 02138, USA}

\author{Ben Falk}
\affiliation{Space Telescope Science Institute, 3700 San Martin Drive, Baltimore, MD, 21218, USA}

\author[0000-0002-2135-9018]{Katharine~M.~Hesse}
\affiliation{Department of Physics and Kavli Institute for Astrophysics and Space Research, Massachusetts Institute of Technology, Cambridge, MA 02139, USA}

\author[0000-0001-9911-7388]{David~W.~Latham}
\affiliation{Center for Astrophysics ${\rm \mid}$ Harvard {\rm \&} Smithsonian, 60 Garden Street, Cambridge, MA 02138, USA}

\author{John Livingston}
\affiliation{Astrobiology Center, 2-21-1 Osawa, Mitaka, Tokyo 181-8588, Japan}
\affiliation{National Astronomical Observatory of Japan, 2-21-1 Osawa, Mitaka, Tokyo 181-8588, Japan}
\affiliation{Astronomical Science Program, Graduate University for Advanced Studies, SOKENDAI, 2-21-1, Osawa, Mitaka, Tokyo, 181-8588, Japan}

\author[0000-0001-7233-7508]{Rachel~A.~Matson}
\affiliation{U.S. Naval Observatory, Washington, D.C. 20392, USA}

\author[0000-0003-0593-1560]{Elisabeth Matthews}
\affiliation{Max-Planck-Institut f\"ur Astronomie, K\"onigstuhl 17, 69117 Heidelberg, Germany}

\author[0000-0003-2058-6662]{George~R.~Ricker}
\affiliation{Department of Physics and Kavli Institute for Astrophysics and Space Research, Massachusetts Institute of Technology, Cambridge, MA 02139, USA}

\author{Alexander~Rudat}
\affiliation{Department of Physics and Kavli Institute for Astrophysics and Space Research, Massachusetts Institute of Technology, Cambridge, MA 02139, USA}

\author[0000-0001-5347-7062]{Joshua~E.~Schlieder}
\affiliation{NASA Goddard Space Flight Center, 8800 Greenbelt Rd, Greenbelt, MD 20771, USA}

\author[0000-0002-6892-6948]{S.~Seager}
\affiliation{Department of Physics and Kavli Institute for Astrophysics and Space Research, Massachusetts Institute of Technology, Cambridge, MA 02139, USA}
\affiliation{Department of Earth, Atmospheric and Planetary Sciences, Massachusetts Institute of Technology, Cambridge, MA 02139, USA}
\affiliation{Department of Aeronautics and Astronautics, MIT, 77 Massachusetts Avenue, Cambridge, MA 02139, USA}

\author[0000-0002-4265-047X]{Joshua N.\ Winn}
\affiliation{Department of Astrophysical Sciences, Princeton University, Princeton, NJ 08544, USA}




\begin{abstract}

The Transiting Exoplanet Survey Satellite (TESS) continues to dramatically increase the number of known transiting exoplanets, and is optimal for monitoring bright stars amenable to radial velocity (RV) and atmospheric follow-up observations. TOI-1386 is a solar-type (G5V) star that was detected via TESS photometry to exhibit transit signatures in three sectors with a period of 25.84 days. We conducted follow-up RV observations using Keck/HIRES as part of the TESS-Keck Survey (TKS), collecting 64 RV measurements of TOI-1386 with the HIRES spectrograph over 2.5 years. Our combined fit of the TOI-1386 photometry and RV data confirm the planetary nature of the detected TESS signal, and provide a mass and radius for planet b of $0.148\pm0.019$~$M_J$ and $0.540\pm0.017$~$R_J$, respectively, marking TOI-1386~b as a warm sub-Saturn planet. Our RV data further reveal an additional outer companion, TOI-1386~c, with an estimated orbital period of 227.6~days and a minimum mass of $0.309\pm0.038$~$M_J$. The dynamical modeling of the system shows that the measured system architecture is long-term stable, although there may be substantial eccentricity oscillations of the inner planet due to the dynamical influence of the outer planet.
\end{abstract}


\section{Introduction}

The plethora of exoplanet discoveries is primarily a result of the transit and radial velocity (RV) methods, the combination of which account for $\sim$95\% of the currently known inventory, according to the NASA Exoplanet Archive (NEA) \citep{akeson2013}. The Transiting Exoplanet Survey Satellite (TESS) is an all-sky photometric survey searching for transiting planets around the nearest and brightest stars \citep{ricker2015}. At the time of writing there are 415 confirmed TESS planets, with 7027 candidates awaiting confirmation\footnote{As of Jan 16, 2024}. 
The observational resources required for follow-up RV observations are significant \citep{kane2009c,cloutier2018,dalba2019c,dragomir2020,guerrero2021,kane2021b} and a coordinated RV campaign has been undertaken since TESS launched in 2018. The TESS-Keck Survey (TKS) is a multi-institutional collaboration focused on planetary occurrence rates, formation, evolution, and dynamics \citep{Chontos2022} and has directly confirmed numerous TESS candidates through precise mass measurements \citep{dalba2020a,Weiss2021,Dai2020,Rubenzahl2021,Scarsdale2021,MacDougall2021,Dalba2022,Lubin2022,Dai2021,Turtelboom2022}. TKS primarily utilises the HIRES spectrometer on the Keck I Telescope at the W.M. Keck Observatory on Maunakea \citep{vogt1994} and the Levy spectrometer on the Automated Planet Finder (APF) \citep{radovan2014,vogt2014a} to confirm and characterize TESS objects of interest (TOIs). As TESS continues to survey the sky and TESS planets are confirmed, the planets discovered will provide answers to some of the remaining questions regarding demographics of planet populations. By surveying our nearest bright stars, the planets discovered will give a greater understanding of planet populations for a variety of stars compared to other surveys, such as Kepler, which focused on sun-like stars \citep{borucki2010a}. This will allow for validation (or invalidation) of the demographics found in the planet populations across all stellar types, such as the existence of the sub-Saturn valley \citep{Ida2004} and as a function of galactic latitude \citep{Zink2023}.

TOI-1386 (TIC 343019899) is a relatively bright ($V = 10.51$) solar-type star, for which a transit signal was detected during Sector 16 of TESS observations. This planet, designated TOI-1386~b, was adopted into the TKS single-transit program, and subsequent RV monitoring of the star commenced soon afterward. Early modeling of the transit and RV data indicated a $\sim$30~day orbital period for the planet, and a size and mass that places it within the category of a warm sub-Saturn. Such planets are an important component of exoplanet demographical studies, and can fall within important gaps in the known exoplanet population \citep{ford2014,winn2015,fulton2017,dulz2020,Hill2023,ostberg2023}. Study of the system was further motivated by the prospect of additional, non-transiting, giant planets that may lie in the system, the demographics of which can inform the early migration history and the influence on planetary architectures \citep{trilling1998,alibert2005,bonomo2017}. Additional RV monitoring of the system successfully revealed another giant planet in the system, designated TOI-1386~c, whose orbital period of $\sim$227~days and moderate eccentricity creates an interesting dynamical environment. Multi-planet giant systems, such as TOI-1386, continue to contribute toward the inventory of exoplanet architectures that diverge significantly from that seen in the solar system \citep{horner2020b,kane2021d,mishra2023a,mishra2023b}.

Here, we report on a detailed analysis of new measurements for the TOI-1386 system, confirming the planetary nature of the transit signals for the inner planet, detecting the presence of a non-transiting outer planet, and measuring the masses and orbital characteristics for both planets. Section~\ref{datasources} details the various data sources used in this analysis, including TESS photometry, high resolution imaging, and Keck RVs. Section~\ref{star} describes the properties of the host star, based on a combination of literature data and analysis of Keck spectra. Section~\ref{Model} presents the results of our data modeling, including the Keplerian orbits for the two discovered planets and a dynamical simulation of their gravitational interactions. Section~\ref{Disc} discusses the implications of the results, particularly for understanding exoplanet demographics in the sub-Saturn valley, and we provide concluding remarks in Section~\ref{Concl}.


\section{Observations}
\label{datasources}


\subsection{Photometry}
\label{dataPHOT}

\begin{figure*}
\centering 
\subfloat{%
  \includegraphics[width=1.8\columnwidth]{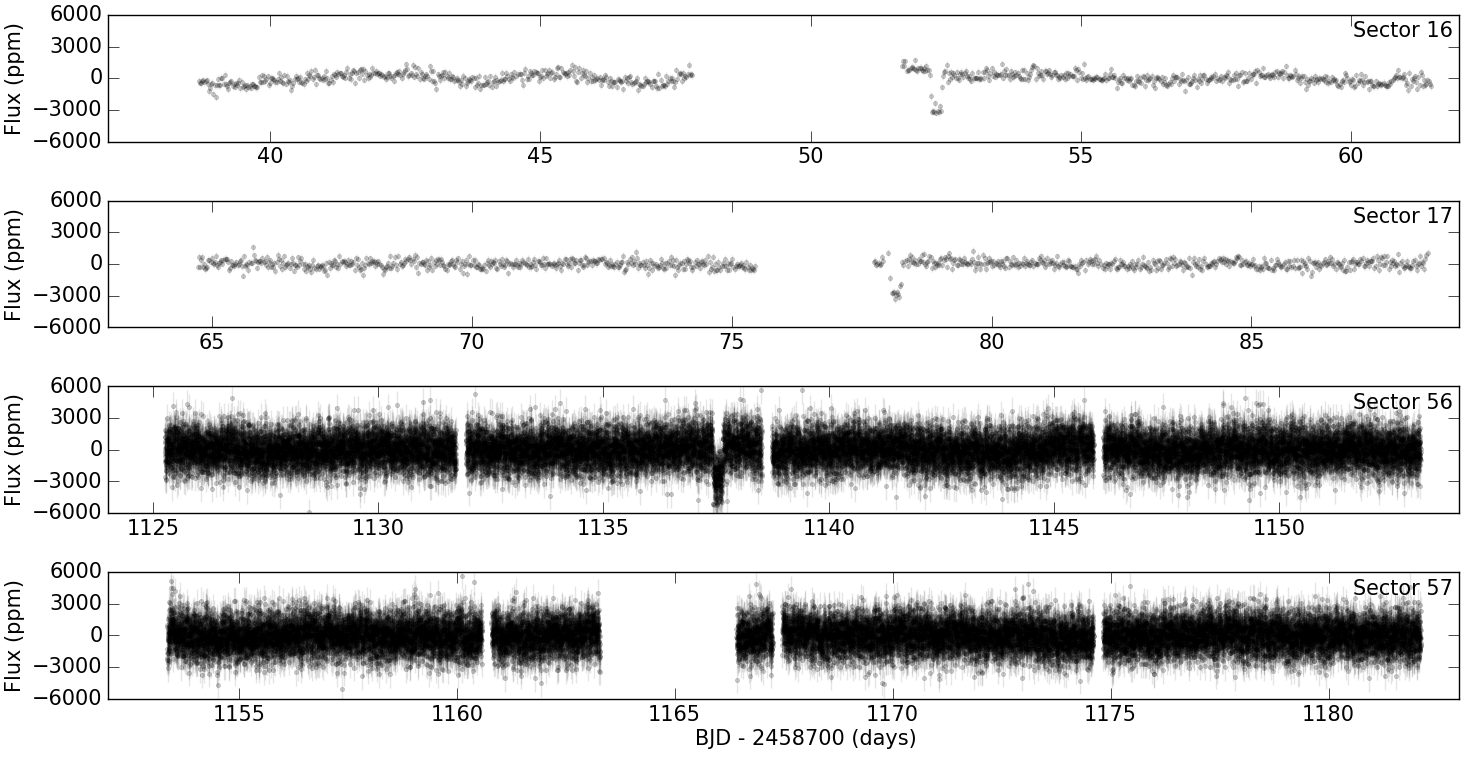}%
}

\subfloat{%
  \includegraphics[width=0.9\columnwidth]{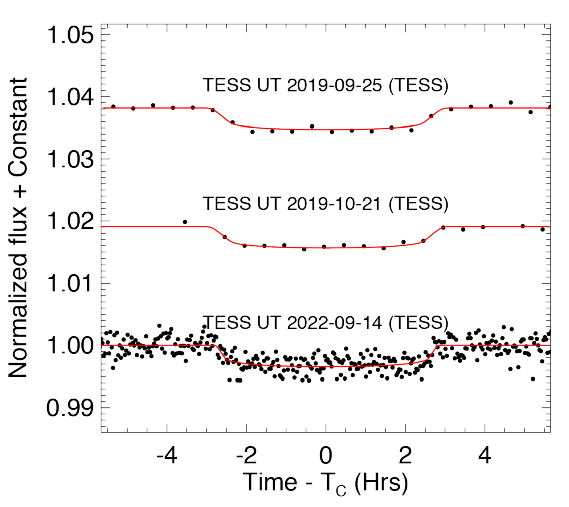}%
}
\subfloat{%
  \includegraphics[width=0.9\columnwidth]{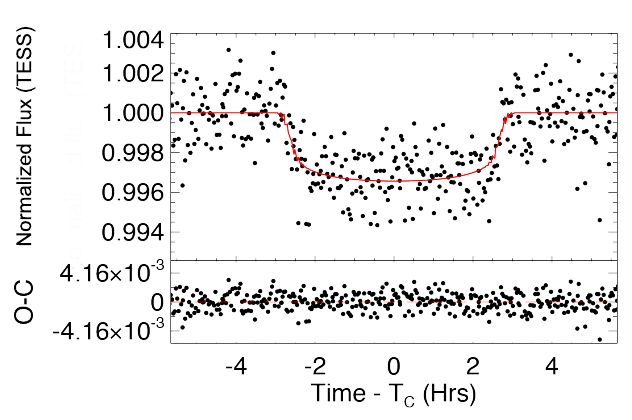}%
  }
\caption{Top: TESS full lightcurve for Sectors 16, 17, 56 and 57. Sectors 16 and 17 are 30-min cadence, processed from the FFIs and sectors 56 and 57 are 2-min cadence processed by the SPOC pipeline. TESS discovered 3 transits of TOI-1386 in (from top) Sectors 16, 17 and 56. The best fit period for the transiting planet positions a transit in the data gap of Sector 57 and hence a 4th transit was not recovered.
Bottom Left: The 3 transits of TOI-1386 discovered by TESS in (from top) Sectors 16, 17 and 56. The red lines show the best fit model to each individual transit by \texttt{EXOFASTv2}. Bottom Right: \texttt{EXOFASTv2} combined these transits with the RV data to determine} the best fit model (red line). The residuals to the fit are shown below the transit model. 
\label{fig:TESS}
\end{figure*}

TESS observed TOI-1386 at a 30-minute cadence in Sectors 16 and 17 from 2019-Sep-11 to 2019-Nov-02, and then again at 2-minute cadence in Sectors 56 and 57 from 2022-Sep-01 to 2022-Oct-29. Thus, the star was observed for a total of 4 Sectors, providing photometric data over a period of $\sim3.1$ years. We initially downloaded data processed through the Quick Look Pipeline \citep[QLP;][]{huang2020a,huang2020b} via the Mikulski Archive for Space Telescopes (MAST) portal \citep{MAST}.\footnote{The data described here may be obtained from the MAST archive at \dataset[doi:10.17909/fwdt-2x66]{https://dx.doi.org/10.17909/fwdt-2x66}.}
Only a single transit event was initially detected in Sector 16 through the QLP, though when looking at the raw data a second transit was detected at the beginning of Sector 17. We accessed the Pre-search Data Conditioning Simple Aperture Photometry \citep[PDC-SAP;][]{stumpe2012,smith2012d} through MAST, cleaned the TESS photometry by keeping only points with quality\_flag = 0, indicating that there is no known problem or condition with the data, and
stitched together and detrended the light curves from TESS sectors 16, 17, 56, and 57 into a single time-series using the Lightkurve software package
\citep{lightkurve}.  
A single transit each was detected in TESS sectors 16, 17, and 56, resulting in an estimated orbital period for TOI-1386~b of $\sim$26~days. No transits were detected in Sector 57, as the predicted transit falls within the data gap for that sector. The top four panels of Figure~\ref{fig:TESS} show the TESS photometry for the sectors during which the star was observed. The bottom panels show the detected transit signatures (bottom left) and folded transit (bottom right), where the red line indicates the best fit model to the transit.

\setlength{\belowcaptionskip}{2pt}


\subsection{Adaptive Optics and Speckle Imaging}
\label{dataAO}

\begin{figure*}
\centering 
\subfloat{%
  \includegraphics[width=1.0\columnwidth]{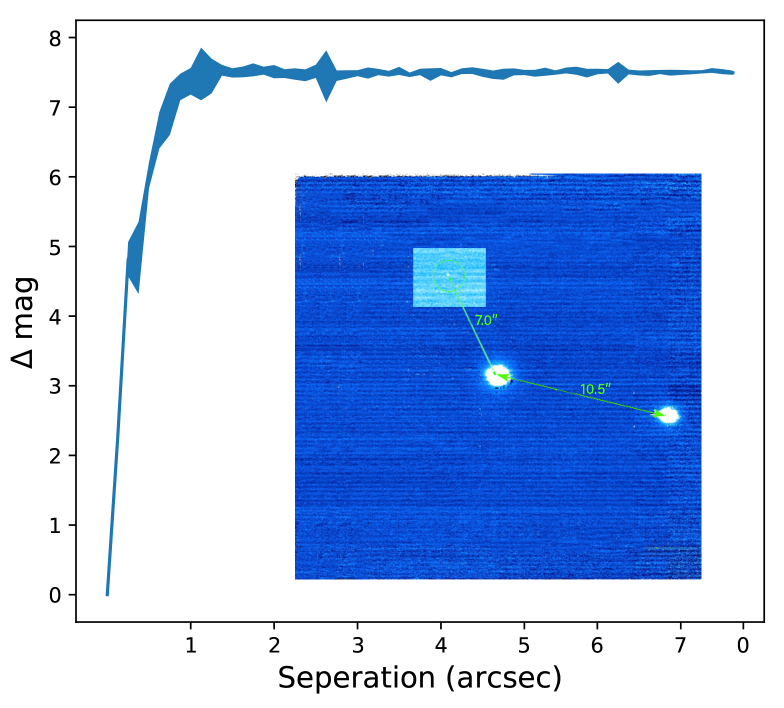}%
}
\subfloat{%
  \includegraphics[width=1.08\columnwidth]{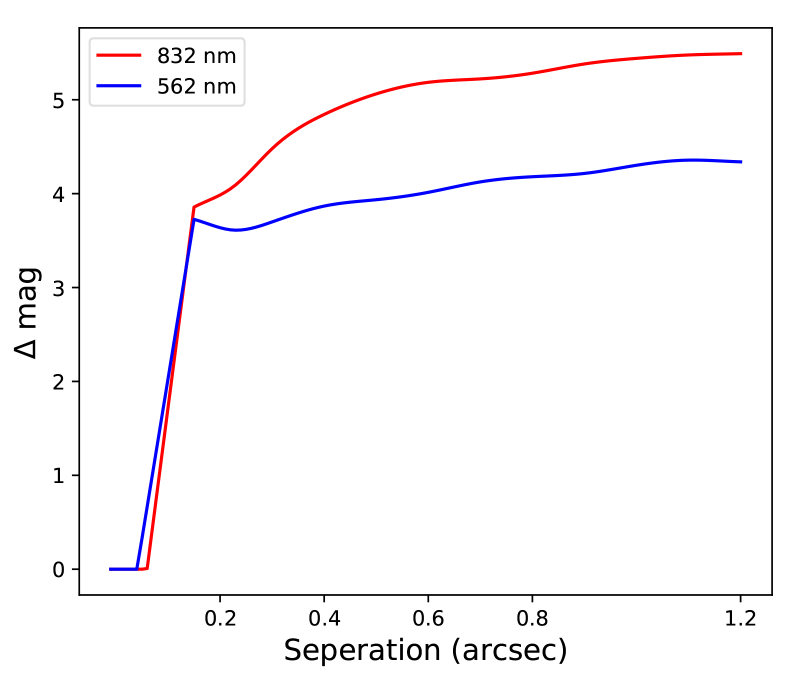}%
  }
\caption{Left: The Gemini/NIRI Contrast curve and AO image (inset). The curve shows the limiting magnitudes ($\Delta mag$) for the non-detection of a neighboring star. At 1" of separation we achieve a $\Delta mag >7$. Two visual neighbors can be seen in the AO image, one with a magnitude difference of $\Delta$~K~mag$=1.25\pm0.01$ and separation of $\sim$10.5\arcsec. The other has a separation of $\sim$7\arcsec\ and a $\Delta K mag = 7.23\pm0.13$ (Shown with a different color scaling). Both of these neighboring stars have a parallax difference to TOI~1386 of $\geq~7~mas$ \citep{GaiaDR3_2021}.
 Right: Limiting magnitudes for the nondetection of a neighboring star
based on the speckle imaging from the NESSI
speckle imager on the WIYN Telescope.
\label{fig:speckle}}
\end{figure*}

We acquired adaptive optics (AO) imaging of TOI-1386 to search for nearby neighboring stars that may have affected or diluted the transit signal discovered by TESS. AO observations of TOI-1386 occurred on 8 Nov 2019 with Gemini/NIRI \citep{hodapp2003}\footnote{Gemini/NIRI data are available on ExoFOP, https://exofop.ipac.caltech.edu/tess/}. At 0.2\arcsec, which at a distance of 146.86~pc \citep{GaiaDR3_2021} corresponds to a projected separation of $\sim$~29~AU, we achieve a detection sensitivity of $\Delta \mathrm{mag} > 4$, and at 1\arcsec\ of separation we achieve a $\Delta \mathrm{mag} >7$. A visual neighbor with a magnitude difference of $\Delta K \mathrm{mag} = 1.25\pm0.01$ and separation of $\sim$10.5\arcsec\ can be seen in the AO image (Figure~\ref{fig:speckle}). An additional visual companion was detected with a separation of $\sim$7\arcsec\ with a $\Delta K \mathrm{mag} = 7.23\pm0.13$. Extracting the details of these stars from \citet{GaiaDR3_2021} we find the $\Delta K \mathrm{mag} = 1.25\pm0.01$ companion is a 15.2~$R_\odot$ star at a distance of 2924.1~pc. There are two Gaia sources $\sim$7\arcsec\ from TOI~1386, one is 10 magnitudes fainter in Gaia G magnitude and is likely a background source. The other, with a Gaia G magnitude difference of $\sim$~8, has a parallax difference of 7.6~mas.
Using the method outlined in \citet{Ciardi2015} to compensate for the flux dilution of neighbouring stars, we calculated that the $\Delta K \mathrm{mag} = 1.25\pm0.01$ companion could cause an error in the measurement of the planet radius of 8.7\%. The $\Delta K \mathrm{mag} = 7.23\pm0.13$ companion made only a negligible difference of 0.1\% of the measured radius of the transiting planet.  

Speckle imaging was also acquired with the NESSI speckle imager \citep{Scott2018,Scott2019} mounted on the 3.5~m WIYN Telescope at Kitt Peak National Observatory on November 17, 2019. NESSI simultaneously acquires data in two bands centered at 562\,nm and 832\,nm using high speed electron-multiplying CCDs (EMCCDs). We collected and reduced the data following the procedures described in \citet{Howell2011}. We were able to determine the limiting magnitude difference ($\Delta$mag) in both the blue (562~nm) and red (832~nm) band as a function of angular separation and found the speckle imaging from WIYN confirmed that no companions are visible from 0.05 -- 1.2\arcsec. Both the AO and speckle images rule out the possibility that the transit signal observed by TESS was caused by a blended stellar companion or eclipsing binary down to the detection limit of the imaging data.

\subsection{Radial Velocities}
\label{dataRV}

We collected 64 spectra of TOI-1386 with the High Resolution Echelle Spectrometer (HIRES) instrument at the Keck Observatory \citep{vogt1994} between UT 2019-12-15 and UT 2022-06-11, including one high signal-noise-ratio (SNR) iodine-free template spectrum. The remaining observations included a gaseous iodine cell which was placed in the light path of the spectrometer. This allowed molecular absorption lines to be combined with the observed stellar spectrum. These lines were used as a wavelength reference for measuring the relative Doppler shift of each spectrum taken during our observations \citep{valenti1995a, butler1996}. We reduced the spectra via the procedure outlined by the California Planet Search \citep[CPS;][]{Howard_2010}. The time series RVs are provided in Table~\ref{tab:RVData} and shown in the top panel of Figure~\ref{fig:RV}. Our RVs have a median nightly binned uncertainty of
$1.71~m~s^{-1}$ and a median SNR of $\sim$203 per pixel at the iodine wavelength region of $\sim$500 nm. Our 2.5 years of observations found both the transiting planet signal of 25.8 days along with an additional longer period, 227.6 day eccentric ($e = 0.27$) Keplerian signal. We also detect a long-term acceleration of the star. These discoveries are discussed further in Sections \ref{ModelRV} and \ref{ModelEXOFAST}.

\begin{deluxetable*}{lccccc}
\tablecaption{\label{tab:RVData} Radial Velocity Time Series}
\tablehead{\colhead{BJD$_{TBD}$} & \colhead{RV (m/s)} & \colhead{RV err (m/s)} & \colhead{S-Value} & \colhead{S-Value err} & \colhead{Instrument}}
\startdata
2458832.865373	& -26.32	& 2.32	& 0.1455	& 0.0010	& HIRES \\
2458852.737299	& -9.61	    & 1.482	& 0.148	    & 0.0010	& HIRES \\
2458869.760972	& -21.53	& 1.72	& 0.1432	& 0.0010	& HIRES \\
2458878.734780	& -12.07	& 1.77	& 0.1509	& 0.0010	& HIRES \\
2459012.112540	& -16.51	& 2.01	& 0.1490	& 0.0010	& HIRES 
\enddata
\tablecomments{{The full data set in a  machine readable format is available online.}}
\end{deluxetable*}


\section{Host Star Properties}
\label{star}

We analyzed our high S/N iodine-free HIRES spectrum of TOI-1386 with the \texttt{SpecMatch} code \citep{CKS1} to derive the $T_{\mathrm{eff}}$ (5828.39$\pm$100.00~K), $\log g$ (4.39$\pm$0.10~cm~s$^{-2}$), and metallicity $[\mathrm{Fe}/{\rm{H}}]$ (0.16$\pm$0.06~dex) of the host star. These results were used as input parameters for the \texttt{EXOFASTv2} fit described in this section and in Section \ref{ModelEXOFAST}. 

We derived the stellar mass, radius, and age using the \texttt{isoclassify} package \citep{Huber2017, Berger2020}. The analysis by \texttt{isoclassify} incorporated Gaia DR2 parallaxes \citep{GaiaCollab}, 2MASS apparent $K$ magnitude, the MESA Isochrones and Stellar Tracks models \citep[MIST;][]{Choi2016} along with priors from the preferred \texttt{SpecMatch} model to measure the best fitting solution of stellar properties. 

We used \texttt{EXOFASTv2} to model the spectral energy distribution (SED) of the star together with the Gaia EDR3 parallax in order to measure the stellar radius. \texttt{EXOFASTv2} computes
SEDs from MIST models and Gaia parallax measurements and fits them to archival photometry to infer the properties of the host star. We used archival broadband photometry of TOI-1386 from Gaia DR2.
We placed a upper bound on extinction ($A_V$ $\leq$ 1.841) from the galactic dust maps of \citet{schlafly2011}. 
We used the \texttt{SpecMatch} analysis of TOI-1386 for the $[\mathrm{Fe}/{\rm{H}}]$ and $T_{\mathrm{eff}}$ priors. This provided a calculated stellar radius of $R_{*} = 1.015^{+0.028}_{-0.026}$~${R}_{\odot }$. The values derived by \texttt{isoclassify} agree with those from \texttt{EXOFASTv2} fit which are provided in Table \ref{tab:TOI1386_star.} 

Reconnaissance spectra of TOI-1386 was taken on December 4, 2021 UT and December 5, 2021 UT with the Tillinghast
Reflector Echelle Spectrograph \citep[TRES;][]{Furesz2008} located at the Fred Lawrence Whipple Observatory (FLWO) in Arizona, USA. We compared the stellar parameters based on the Stellar Parameter Classification \citep[SPC;][]{buchhave2012} analysis of six TRES spectra with our results from \texttt{EXOFASTv2} and found the values agreed within the uncertainties.

TOI-1386 is a chromospherically inactive star with $\log R^{\prime}_{HK}$ = -5.00 $\pm$ 0.05. Analysis of the TESS photometry found no significant periodicities that would imply the transit signal was caused by stellar activity. We computed the Ca II H\&K index ($S_{HK}$) as described in \citet{IandF2010} for our Keck/HIRES time series data (Table \ref{tab:RVData}). We found there was a correlation between the $S_{HK}$ values and RVs. Section \ref{test} outlines our methods to test the robustness of the RV planet detection and rule out stellar activity as the cause of the RV signal.


\section{Data Modeling}
\label{Model}


\subsection{Initial Spectroscopy Model with \texttt{RadVel}}
\label{ModelRV}

The RV data were initially fit using the RV modeling toolkit \texttt{RadVel} \citep{fulton2018a}. 
\texttt{RadVel} enables users to model Keplerian orbits in radial velocity time series. \texttt{RadVel} uses an iterative approach to determine the best-fit orbital parameters for the observed RV curve. It then employs modern Markov chain Monte Carlo (MCMC) sampling techniques \citep{metropolis53,hastings70,Foreman2013} to infer orbital parameters along with their associated uncertainties.

The RV data was fit with \texttt{RadVel} in order to model the $\sim$28~day period that was predicted from the initial single transit duration via the method outlined in \citet{Seager2010}. Early fits detected both the transiting planet period of 25.8~days along with a linear trend, which later resolved into the outer planet period of 227.6~days. The final preferred fit with \texttt{RadVel} included both planets b~\&~c and a 4-sigma linear trend. This preferred model was confirmed in the final combined model fit we completed with \texttt{EXOFASTv2} (Table \ref{tab:TOI1386_planet.} and Figure \ref{fig:RV}).


\subsection{Testing the RV Model}
\label{test}

As mentioned in Section~\ref{star}, we found a Pearson correlation coefficient of 0.62 between the $S_{HK}$ values and RVs, with the S-value periodogram peaking at 234 days. To determine if the outer planet detection was robust we conducted a series of tests that would rule out stellar activity as the cause of the signal. We decorrelated the RV data by fitting a line of best fit to the S-values, removing this signal from the RVs and remodeling our best fit solution. We fit the decorrelated RVs and found the best fit had negligible changes from the original.
We then added a planet to the RV fit with the period set to be equal to the value of the peak in the S-values and ran the model again. The best fit parameters for TOI-1386~b and TOI-1386~c did not change. We also ruled out sidereal day, lunation period, and sidereal year aliases as potential causes of the RV signal. We conclude from these tests that the RV signals are not an alias, nor are they caused by stellar activity.

As a typical stellar activity cycle is much longer than 234 days and stellar rotation rates tend to be shorter, we are unsure as to the cause of this peak in the s-values. We encourage further study into the possible causes of a peak at this period.

\begin{figure*}
\centering 
\subfloat{%
\includegraphics[width=1.5\columnwidth]{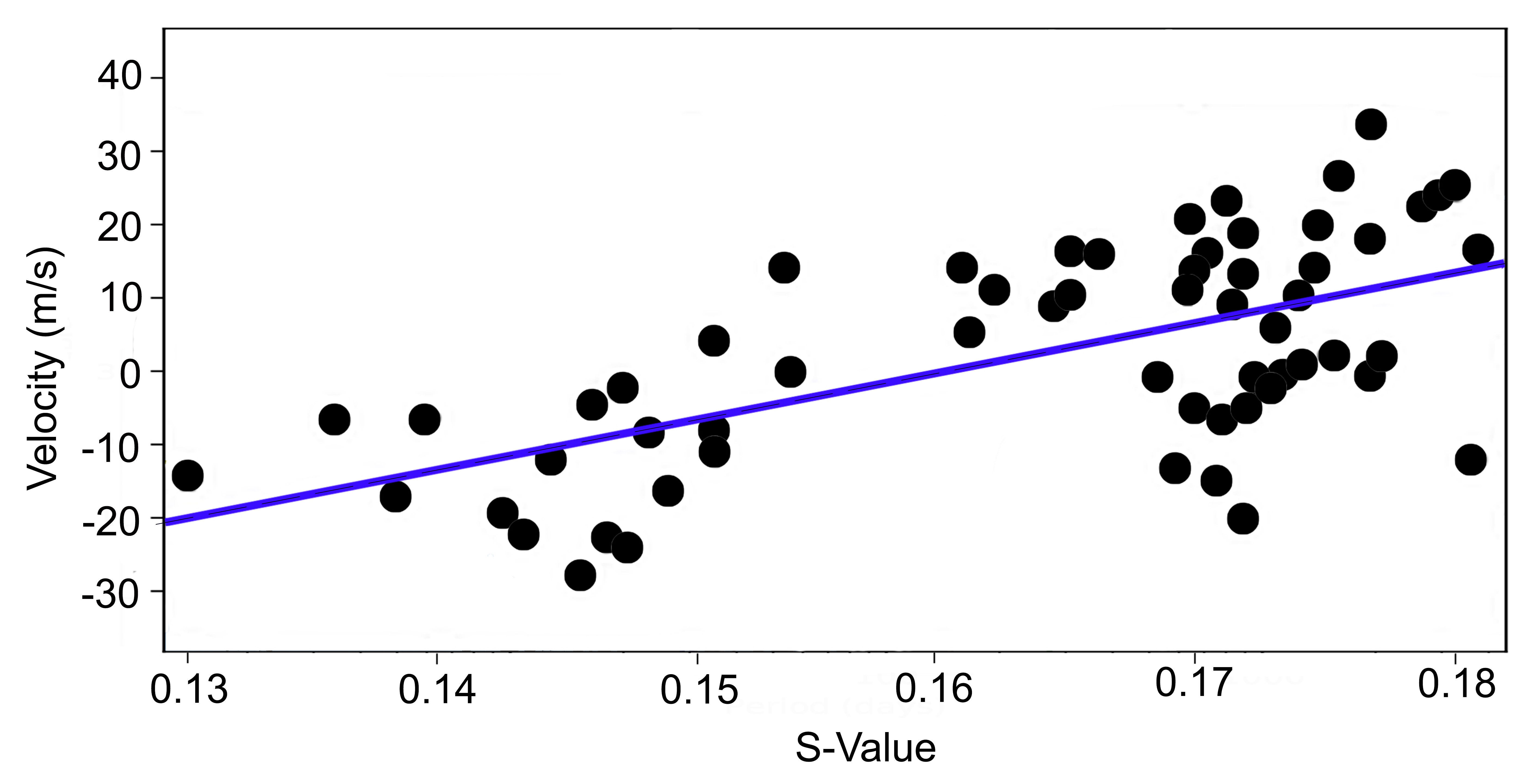}%
}
\quad
\subfloat{%
\includegraphics[width=2.0\columnwidth]{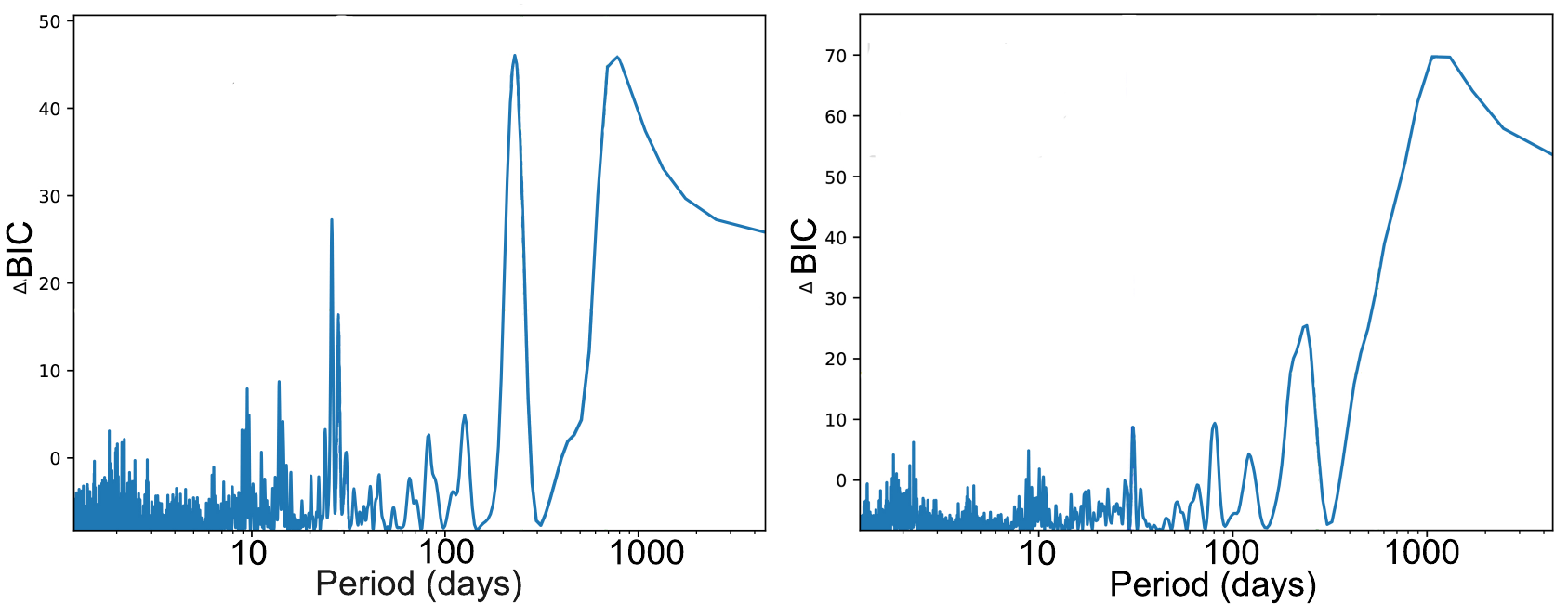}%
  }
\caption{ Top: The S-Value and Velocity correlation. We found a Pearson
correlation coefficient of 0.62 between the $S_{HK}$ values and RVs. Bottom Left: Radial Velocity periodogram, with a maximum peak at 227 days. Note the secondary peak at 787 days was investigated and ruled out as the solution to the fit. Bottom Right: The S-value periodogram, with a maximum peak at $\sim 1300$ days and another peak at 234 days. We tested our model for the outer planet via the methods outlined in Section \ref{test} and concluded that the RV signals were not caused by stellar activity.} 
\label{fig:svalues}
\end{figure*}


\subsection{Combined System Modeling with \texttt{EXOFASTv2}}
\label{ModelEXOFAST}

We modeled the stellar and planetary parameters for the TOI-1386 system using the \texttt{EXOFASTv2} modeling suite
\citep{eastman2013,eastman2020}. We included the TESS transit photometry along with the RVs from Keck-HIRES. The 30 minute cadence data from Sectors 16 \& 17 with the 120 second cadence data from Sectors 56 \& 57 were detrended using the spline fitting tool \texttt{keplersplinev2} \footnote{https://github.com/avanderburg/keplersplinev2}. 
For the 30-minute cadence we use the oversampling feature of \texttt{EXOFASTv2}, as 30 minutes is long relative to the ingress and egress times ($\tau~\sim$~18~minutes). The oversampling feature samples the model at 10 points between $\pm$~15 minutes of that timestamp, then numerically integrates the model over that range to get the single model point. This is important to avoid misfitting the transit shape. 

\begin{figure*}
\centering 
\subfloat{%
  \includegraphics[width=1.5\columnwidth]{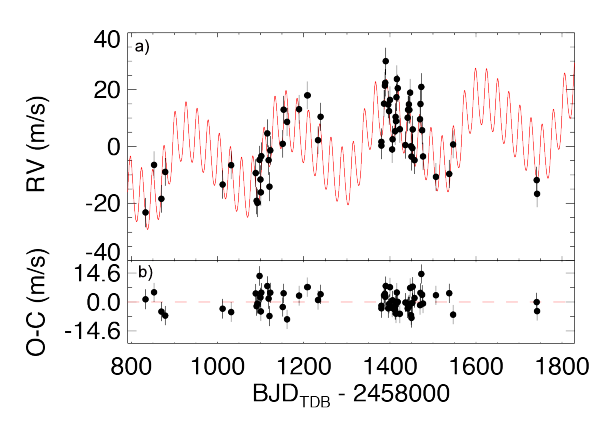}%
}

\subfloat{%
  \includegraphics[width=0.95\columnwidth]{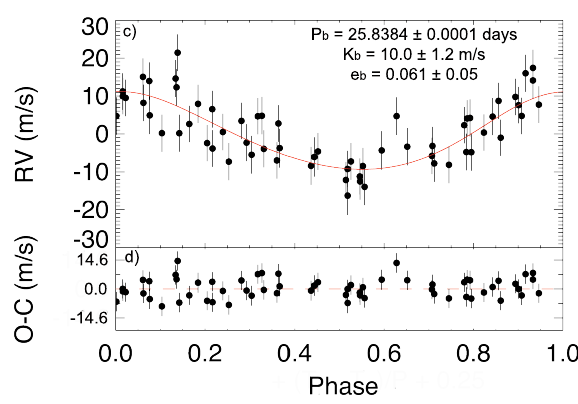}%
}
\subfloat{%
  \includegraphics[width=0.95\columnwidth]{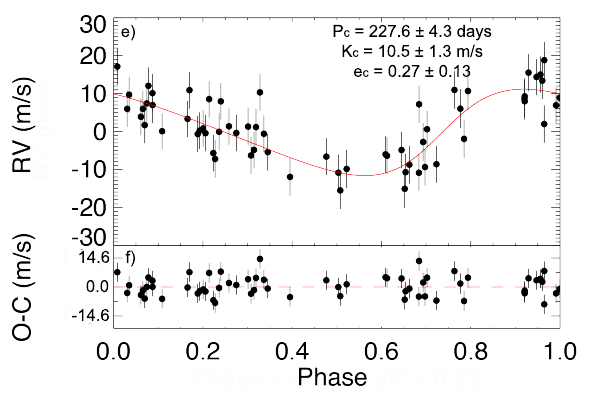}%
  }
\caption{The RV measurements of TOI-1386. Panel a: Our complete RV time series with our preferred model shown in red. Panel b: residuals between the RVs and best-fit 2 planet model. Panels c and e: Phase-folded RV time series for planets b and c respectively. Panels d and f: The residuals to each of the planet fits. 
\label{fig:RV}}
\end{figure*}

When running the joint fit we placed normal priors on $T_{\rm eff}$  (5828.39$\pm~100.00$~K) and $[\mathrm{Fe}/{\rm{H}}]$ (0.16$\pm0.06$~dex) from \texttt{SpecMatch}. The SED model discussed in Section \ref{star} was simultaneously fit along with the transit and RV models. We found the preferred model included 2 planets as well as a 4-sigma linear trend ($\dot{\gamma}$) (Table \ref{tab:TOI1386_star.}). The best-fit transit model is shown in Figure~\ref{fig:TESS} along with the data taken by TESS. Only three transits are depicted, as the transit that was set to occur in Sector 57 coincided with a data gap (Figure \ref{fig:TESS}, bottom left). All RV data and the best-fit RV model are shown in Figure \ref{fig:RV}. The RV and the transit fits agreed, giving the interior planet, TOI-1386~b,  a period of 25.8384 days and eccentricity of 0.061 (consistent with 0). The transiting planet has a radius of 0.540~$R_J$ and a measured mass of 0.148~$M_J$, equating to a bulk density of 1.16~g~cm$^{-3}$. This places TOI-1386~b in the sub-Saturn regime. The outer non-transiting planet, TOI-1386~c, was fit with a 227.6 day period and eccentricity of 0.27. This planet has a minimum mass ($M_p\sin~i$) of $0.309~M_J$, making it a warm Saturn mass planet with an equilibrium temperature ($T_{eq}$) of 327.4~K (assuming no albedo and perfect temperature redistribution). The preferred fit also included a 4-sigma linear trend ($\dot{\gamma}$) which is an indication of an additional orbiting body, and further observations of TOI-1386 are needed to determine if the cause of this trend is due to an outer sub-stellar companion. The values included in this discussion and their uncertainties, along with many additional derived stellar and planet parameters provided by \texttt{EXOFASTv2}, are listed in Tables \ref{tab:TOI1386_star.} and \ref{tab:TOI1386_planet.}.

\startlongtable
\begin{deluxetable*}{lcccccccc}
\tablecaption{Stellar Parameters}
\tablehead{\colhead{~~~Parameter} & \colhead{Units} & \colhead{Values} & \colhead{Source}}
\startdata
~~~~$RA$\dotfill &Right Ascension\dotfill &22:18:00.98 & \citet{GaiaCollab}\\
~~~~$DEC$\dotfill & Declination\dotfill &  +54:19:08.00 & \citet{GaiaCollab} \\
~~~~$V mag$\dotfill &V Magnitude (mag)\dotfill &  10.51 & \citet{GaiaCollab} \\
~~~~$d$\dotfill & Distance (PC)\dotfill &  $146.858 \pm 0.684$ & \citet{GaiaCollab} \\
~~~~$Type$\dotfill & Spectral Type\dotfill & G5V & \citet{Simbad} \\
~~~~$M_*$\dotfill &Mass (\msun)\dotfill &$1.038^{+0.050}_{-0.058}$ & This work\\
~~~~$R_*$\dotfill &Radius (\rsun)\dotfill &$1.015^{+0.028}_{-0.026}$ & This work\\
~~~~$L_*$\dotfill &Luminosity (\lsun)\dotfill &$1.039^{+0.064}_{-0.042}$ & This work\\
~~~~$\rho_*$\dotfill &Density (cgs)\dotfill &$1.40\pm0.13$ & This work\\
~~~~$\log{g}$\dotfill &Surface gravity (cgs)\dotfill &$4.441^{+0.028}_{-0.033}$ & This work\\
~~~~$T_{\rm eff}$\dotfill &Effective Temperature (K)\dotfill &$5793^{+76}_{-73}$ & This work\\
~~~~$[{\rm Fe/H}]$\dotfill &Metallicity (dex)\dotfill &$0.164\pm0.056$ & This work\\
~~~~$Age$\dotfill &Age (Gyr)\dotfill &$3.3^{+3.5}_{-2.2}$ & This work\\
~~~~$\varpi$\dotfill &Parallax (mas)\dotfill &$6.768^{+0.032}_{-0.031}$ & \citet{GaiaCollab} \\
~~~~$\dot{\gamma}$\dotfill &RV slope$^{a}$ (m/s/day)\dotfill &$0.0180\pm0.0044$ & This work\\
\enddata
\label{tab:TOI1386_star.}
\tablenotetext{}{ Parameters noted as from ``This work" were derived using \texttt{EXOFASTv2}. See Table 3 in \citet{eastman2020} for a detailed description of all parameters}
\tablenotetext{a}{Reference epoch = 2459287.478334}
\end{deluxetable*}

\startlongtable
\begin{deluxetable*}{lccccccc}
\tablecaption{Planetary Parameters}
\tablehead{\colhead{~~~Parameter} & \colhead{Units} & \multicolumn{6}{c}{Values}}
\startdata
\\\multicolumn{2}{l}{ }&b&c\smallskip\\
~~~~$P$\dotfill &Period (days)\dotfill &$25.83839\pm0.00013$&$227.6^{+4.6}_{-4.0}$\\
~~~~$R_p$\dotfill &Radius (\rj)\dotfill &$0.540^{+0.018}_{-0.016}$&
--\\
~~~~$M_p$\dotfill &Mass (\mj)\dotfill &$0.148^{+0.019}_{-0.018}$&--
\\
~~~~$T_c$\dotfill &Time of conjunction$^{a}$ (\bjdtdb)\dotfill &$2458752.3213^{+0.0033}_{-0.0034}$&$2458771^{+16}_{-18}$\\
~~~~$a$\dotfill &Semi-major axis (AU)\dotfill &$0.1732^{+0.0027}_{-0.0033}$&$0.739^{+0.015}_{-0.016}$\\
~~~~$i$\dotfill &Inclination (Degrees)\dotfill &$89.69^{+0.21}_{-0.25}$&--
\\
~~~~$e$\dotfill &Eccentricity \dotfill &$0.061^{+0.056}_{-0.042}$&$0.27\pm0.13$\\
~~~~$\omega_*$\dotfill &Argument of Periastron (Degrees)\dotfill &$-33^{+48}_{-44}$&$-90^{+27}_{-29}$\\
~~~~$T_{eq}$\dotfill &Equilibrium temperature$^{b}$ (K)\dotfill &$676.4^{+10.}_{-8.3}$&$327.4^{+5.3}_{-4.6}$\\
~~~~$K$\dotfill &RV semi-amplitude (m/s)\dotfill &$10.0\pm1.2$&$10.5\pm1.3$\\
~~~~$\delta$\dotfill &$\left(R_P/R_*\right)^2$ \dotfill &$0.002983^{+0.000090}_{-0.000083}$&--
\\
~~~~$\tau$\dotfill &Ingress/egress transit duration (days)\dotfill &$0.01271^{+0.0016}_{-0.00052}$&--
\\
~~~~$T_{14}$\dotfill &Total transit duration (days)\dotfill &$0.2363^{+0.0028}_{-0.0026}$&--
\\
~~~~$b$\dotfill &Transit Impact parameter \dotfill &$0.20^{+0.17}_{-0.14}$&--\\
~~~~$\rho_p$\dotfill &Density (cgs)\dotfill &$1.16^{+0.19}_{-0.18}$&--\\
~~~~$\log g_p$\dotfill &Surface gravity \dotfill &$3.099^{+0.058}_{-0.064}$&--\\
~~~~$\fave$\dotfill &Incident Flux (F$\oplus$)\dotfill &$34.680^{+2.131}_{-1.763}$&$1.771^{+0.147}_{-0.154}$\\
~~~~$T_P$\dotfill &Time of Periastron (\bjdtdb)\dotfill &$2458744.1^{+3.4}_{-3.5}$&$2458656^{+15}_{-18}$\\
~~~~$e\cos{\omega_*}$\dotfill & \dotfill &$0.041^{+0.048}_{-0.040}$&$-0.00^{+0.12}_{-0.11}$\\
~~~~$e\sin{\omega_*}$\dotfill & \dotfill &$-0.022^{+0.030}_{-0.054}$&$-0.25\pm0.13$\\
~~~~$M_p\sin i$\dotfill &Minimum mass (\mj)\dotfill &$0.148^{+0.019}_{-0.018}$&$0.309\pm0.038$\\
~~~~$M_p/M_*$\dotfill &Mass ratio \dotfill &$0.000137\pm0.000016$&$0.000350^{+0.00036}_{-0.000065}$\\
\smallskip\\\multicolumn{2}{l}{Wavelength Parameters:}&TESS\smallskip\\
~~~~$u_{1}$\dotfill &linear limb-darkening coeff \dotfill &$0.294\pm0.030$\\
~~~~$u_{2}$\dotfill &quadratic limb-darkening coeff \dotfill &$0.270\pm0.029$\\
\smallskip\\\multicolumn{2}{l}{Telescope Parameters:}&HIRES\smallskip\\
~~~~$\gamma_{\rm rel}$\dotfill &Relative RV Offset$^{c}$ (m/s)\dotfill &$-2.38^{+0.97}_{-0.99}$\\
~~~~$\sigma_J$\dotfill &RV Jitter (m/s)\dotfill &$5.26^{+0.62}_{-0.54}$\\
~~~~$\sigma_J^2$\dotfill &RV Jitter Variance \dotfill &$27.6^{+7.0}_{-5.4}$\\
\enddata
\label{tab:TOI1386_planet.}
\tablenotetext{}{See Table 3 in \citet{eastman2020} for a detailed description of all parameters}
\tablenotetext{a}{Time of conjunction is commonly reported as the "transit time"}
\tablenotetext{b}{Assumes no albedo and perfect redistribution}
\tablenotetext{c}{Reference epoch = 2459287.478334}
\end{deluxetable*}


\subsection{System Dynamics}
\label{dynamics}

The orbital eccentricity of planet c raises interest concerning the dynamical history of the system. To investigate this, we conducted a dynamical analysis using the N-body simulation package \texttt{REBOUND} \citep{rein2012a} with the symplectic integrator \texttt{WHFast} \citep{Rein2015}. Planetary orbital parameters from Table~\ref{tab:TOI1386_planet.} are taken as the starting condition of the simulation with the assumption that both planets are coplanar. Note that mutual inclinations between planets can result in non-linear trends regarding stable locations within the system, particularly for cases of mean motion resonance \citep{Barnes2015}, but can also resolve cases of orbital instability that result from coplanar assumptions \citep{kane2016d}. Given the age of the system, the mutual inclinations present within this system would present a useful case study on the long-term stability of non-coplanar orbital configurations.

\begin{figure*}
  \begin{center}
    \includegraphics[trim=0 0 0 30,clip,width=1.0\textwidth]{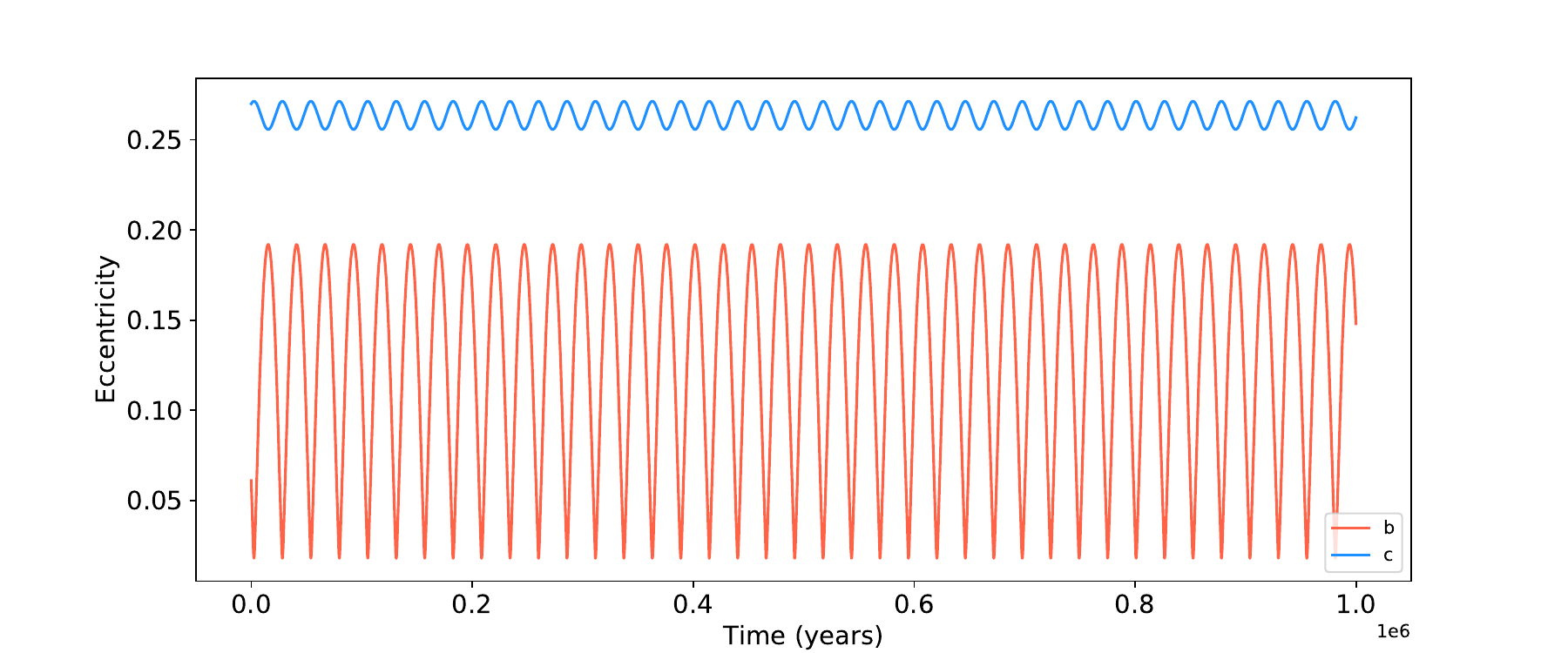}
  \end{center}
  \caption{Eccentricity evolution of the TOI-1386 b and c planets. Shown here are the first $10^6$~years from the full $10^7$~year dynamical simulation, described in Section~\ref{dynamics}.}
  \label{fig:dynamics}
\end{figure*}

Under the assumption that both planets are coplanar, the minimum mass derived for planet c can be considered to be its true mass for the purpose of this study.The simulation was integrated for $10^7$ years with a time step size of 1/20 of the planet b orbital period, or $\sim$1.3~days, in accordance with the recommended time step from \citet{duncan1998} to ensure proper orbital sampling. We recorded the results every 100 years and show the time series eccentricity evolution of the two planets in Figure~\ref{fig:dynamics}. The system is dynamically stable for the duration of the simulation. However, both planets experience eccentricity variations over time through the transfer of angular momentum \citep{chambers1996,rasio1996,laughlin2001,juric2008,kane2013b,kane2021a}. This is especially the case for the orbit of planet b, which varies from nearly circular to almost 0.2 eccentricity, suggesting that present observations happen to be occurring during a period of low eccentricity for the inner planet. Such large variation appears to be due to the influence of the outer planet, which has double the mass of the inner planet, driving the eccentricity evolution of both planets that exhibit in-sync periodic cycles every $\sim$25,800 years. Such an orbital configuration may remain long-term stable, or produce a planet-planet scattering event that leaves one remaining planet in a highly eccentric orbit \citep{chatterjee2008,carrera2019b,kane2023}. The relatively large separation of the two known planets makes a planet-planet scattering scenario unlikely, but the configuration of these planets and their dynamical interactions may exclude the presence of additional, smaller, planets within the inner part of the planetary system.

\section{Discussion}
\label{Disc}

\begin{figure}
  \begin{center}
  \includegraphics[width=0.5\textwidth]{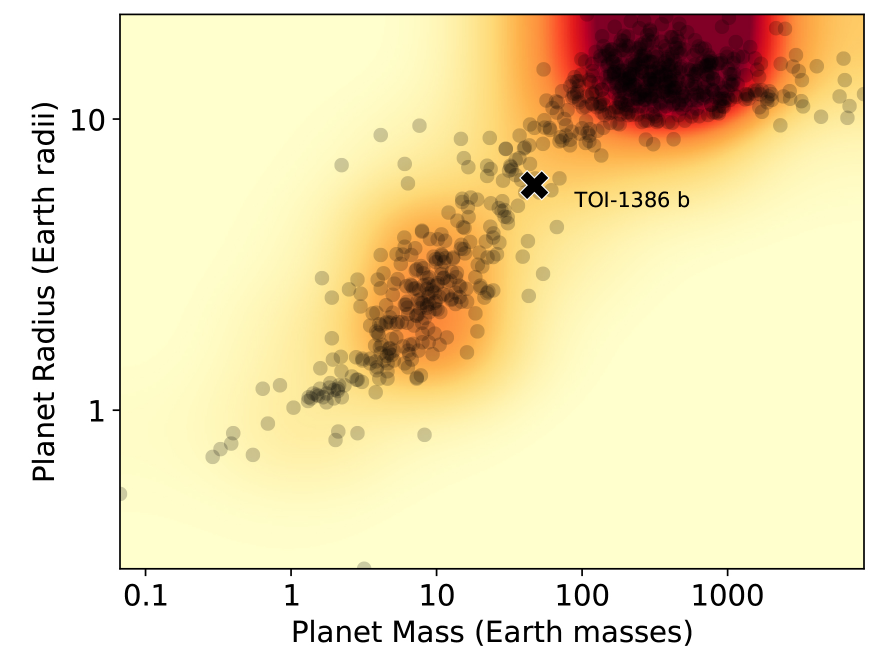}
  \end{center}
  \caption{Heat map of the mass and radius distribution of exoplanets showing the position of TOI-1386~b (black X) compared to the full catalog of known exoplanets (gray). TOI-1386~b is positioned in the sub-Saturn valley between the cluster of Jupiter-sized planets and that of the Super Earths/Neptunes.}
  \label{fig:MR}
\end{figure}

The size and mass of the inner transiting planet place it in the sub-Saturn valley (Figure~\ref{fig:MR}). As described in \citet{Hill2023}, the existence of the sub-Saturn valley is still a topic of debate. Some suggest the sub-Saturn gap could be attributed to core accretion theory \citep{Ida2004, Mayor2011, Emsenhuber2021}, where planets that reach 10~$M_\oplus$ enter a runaway accretion period and rapidly grow to $\geq 100$~$M_\oplus$, provided there are sufficient materials available. Others, including \citet{Suzuki_2018}, argue the apparent gap is merely credited to the difficulty of detecting sub-Saturn mass planets and hence the discovery of TOI-1386~b is an important contribution to this demographic of planets. A study from \citet{Bennett2021} found that the sub-Saturn valley was missing from their analysis of RV planets observed using CORALIE/HARPS. Amongst those Sub-Saturns planets that have been detected there is a large density diversity, with sub-Saturn planet masses ranging from 6 -- 60 $M_\oplus$ regardless of size \citet{petigura2017}. TESS will help to fill out the mass-radius diagram by allowing for RV follow up of many transiting planets around the brightest stars in the sky. The inner transiting planet TOI~1386~b will contribute to the data used to ultimately verify the existence of the sub-Saturn valley and give clues as to the formation processes of these elusive planets.

\begin{figure*}
  \begin{center}
  \includegraphics[width=1.0\textwidth]{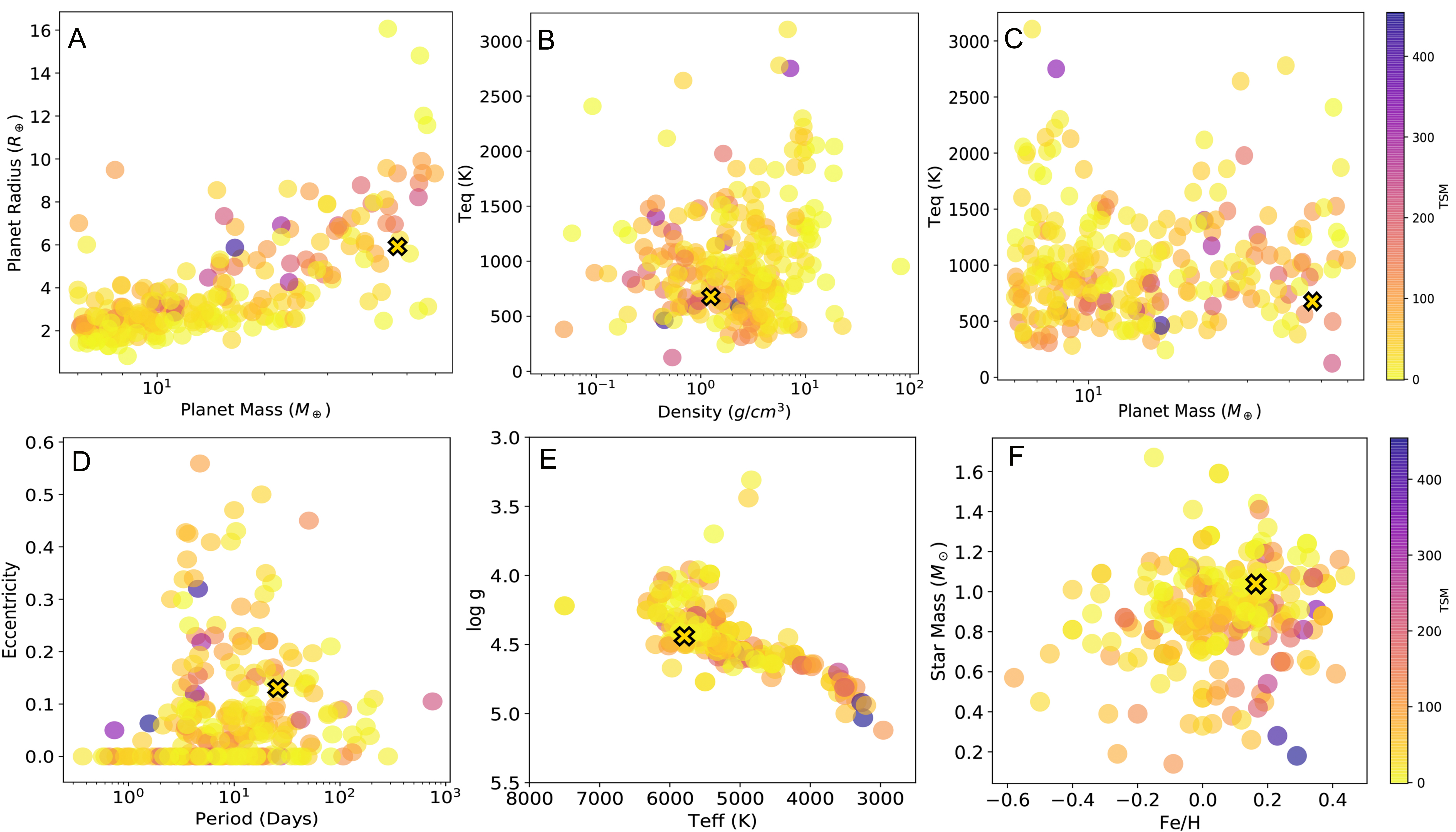}
  \end{center}
\caption{The attributes of planet TOI-1386~b (Panels A--D) and star TOI-1386 (Panels E \& F) compared to the population of sub-Saturn planets. TOI-1386~b's position is indicated by the cross. Planets with a measured mass between 6~--~60~$M_\oplus$ are included in our sub-Saturn population. The colorbar for each plot indicates the TSM value of the planet, with TOI-1386~b having a TSM of 42.5. Compared to the rest of the sub-Saturn population, TOI-1386~b is a massive, cool, moderately eccentric planet orbiting a metal rich sun-like star.
\label{fig:sub-sat}}
\end{figure*}

As TOI-1386~b is a relatively rare sub-Saturn planet we calculated the transmission spectroscopy metric (TSM) value \citep{kempton2018} to determine if it is a good JWST target \citep{gardner06}. The TSM is proportional to the expected transmission spectroscopy signal-to-noise, based on the strength of spectral features, brightness of the host star, and mass and radius of the planet assuming cloud-free atmospheres. We calculate a TSM value of 42.5 for TOI-1386~b. We compared the parameters of this target to the population of known sub-Saturn planets (Figure \ref{fig:sub-sat}). We limit the sub-Saturn valley planets to those with masses between 6 -- 60 $M_\oplus$ in accordance with \citet{petigura2017}. We find that compared to the rest of the sub-Saturn population TOI-1386~b is a massive, cool, moderately eccentric
planet orbiting a metal rich sun-like star. In investigating the uniqueness of TOI-1386~b we found that a large number of detected sub-Saturn planets are a part of multi-planet systems that include multiple sub-Saturn planets within the system. When these rarely found planets are discovered, they are often found in duplicate. Of the 621 sub-Saturns listed on the NEA with a mass between 6-60 $M_\oplus$, there are 400 that are multi-planet systems. Of those systems we found 275 sub-Saturn planets were in systems where there were multiple sub-Saturn 
planets. This indicates that $\sim44\%$ of all sub-Saturn planets detected have additional sub-Saturn planets orbiting in their system. This finding agrees with the \citet{weiss2018a} study that found that in systems with multiple planets, the planets are likely to be of similar sizes. 
In the absence of any correction for detection biases and completeness, we are unable to provide a firm conclusion regarding this finding. More investigation into the sub-Saturn valley population needs to occur to determine whether this is a true phenomena or due to observational biases.

Originally chosen as part of the TKS single transit target list, TOI-1386 is a prime example of the importance of examining TESS raw data \citep{dragomir2020, dalba2020b, Dalba2022}. Investigation of the raw light curves from TESS revealed additional transits that may have initially been missed due to their proximity to data gaps. The discovery of these additional transits plus early RV data collection allowed quick identification of the $\sim26$~day orbit of the transiting planet. Although TOI-1386 was already earmarked for further investigation by the TKS team as a single transit target, the process of scrutinizing the raw data and detecting these extra transits reinforced the likelihood of the candidate being a planet and justified further study.
This may be particularly important for planets that belong to key demographics, such as the sub-Saturn valley. The apparent scarcity of planets in this group may be due to the inherent challenges in detecting them, underscoring the importance of thorough data analysis.


\section{Conclusions}
\label{Concl}

TESS observed TOI-1386 in 4 Sectors (16, 17, 56 $\&$ 57) over a span of 3.2 years. Though initially only a single transit was detected, analysis of TESS data found 3 transits across the 4 Sectors. RV observations of TOI-1386 confirmed the existence of TOI-1386~b with a period of 25.83839 days and gave a mass measurement of $M_p$~=~$0.148~M_J$, placing the $0.540~R_J$ planet in the sub-Saturn regime. A nearby visual neighbor with a separation of $\sim$10.5\arcsec and magnitude difference of $\Delta$~K~mag$=1.25\pm0.01$ was calculated to have a dilution factor that could cause the radius of the planet to be underestimated by 8.7\%. Even with this factor considered TOI-1386~b remains in the sub-Saturn regime and is an important addition to this demographic of planets due to the ongoing debate regarding the origin of the sub-Saturn gap. With a TSM of 42.5, TOI-1386~b is a relatively massive, cool, moderately eccentric
sub-Saturn planet orbiting a metal rich sun-like star.

RV analysis also detected an additional planet signal of an outer, non-transiting planet. TOI-1386~c is a eccentric (0.27) planet with a 227.6 day orbit and a minimum mass of $0.309~M_J$. Even with an eccentric outer planet, dynamical simulations of the system found that the two planets remained stable for the entire simulated period of $10^7$~years. Both planets do, however, experience large changes in eccentricity over the simulation period with the inner planet eccentricity fluctuating from 0 to 0.2 and back every $\sim25,800$~years. The best fit model of TOI-1386 also included a 4-sigma linear trend. Further observations of the star will be required to determine whether the cause of this trend is planetary in nature. 

In our investigation of TOI-1386~b and the sub-Saturn valley population, we found that $\sim44\%$ of all identified sub-Saturn planets are part of systems containing other sub-Saturn planets. This finding suggests a tendency for planets of this size to coexist with similarly sized planets in their orbits. While this statistic lacks correction for detection biases and completeness, we encourage further research into the sub-Saturn valley to determine the validity of this phenomenon.

TOI~1386 is scheduled to be observed again by TESS in Sectors 76, 77 and 83. These observations, along with any additional RV observations, will help to refine the orbital parameters of the TOI~1386 system and determine the cause of the linear trend detected. 

\section*{Acknowledgments}

M.L.H. would like to acknowledge NASA support via the FINESST Planetary Science Division, NASA award number 80NSSC21K1536. E.A.P. acknowledges the support of the Alfred P. Sloan Foundation. L.M.W. is supported by the Beatrice Watson Parrent Fellowship and NASA ADAP Grant 80NSSC19K0597. D.H. acknowledges support from the Alfred P. Sloan Foundation, the National Aeronautics and Space Administration (80NSSC21K0652) and the Australian Research Council (FT200100871). I.J.M.C. acknowledges support from the NSF through grant AST-1824644. P.D. acknowledges support from a 51 Pegasi b Postdoctoral Fellowship from the Heising-Simons Foundation. A.B. is supported by the NSF Graduate Research Fellowship, grant No. DGE 1745301. R.A.R. is supported by the NSF Graduate Research Fellowship, grant No. DGE 1745301. C.D.D. acknowledges the support of the Hellman Family Faculty Fund, the Alfred P. Sloan Foundation, the David \& Lucile Packard Foundation, and the National Aeronautics and Space Administration via the TESS Guest Investigator Program (80NSSC18K1583). J.M.A.M. is supported by the NSF Graduate Research Fellowship, grant No. DGE-1842400. J.M.A.M. also acknowledges the LSSTC Data Science Fellowship Program, which is funded by LSSTC, NSF Cybertraining Grant No. 1829740, the Brinson Foundation, and the Moore Foundation; his participation in the program has benefited this work. T.F. acknowledges support from the University of California President's Postdoctoral Fellowship Program.J.V.Z. acknowledges support from the Future Investigators in NASA Earth and Space Science and Technology (FINESST) grant 80NSSC22K1606.

We thank the time assignment committees of the University of California, the California Institute of Technology, NASA, and the University of Hawaii for supporting the TESS-Keck Survey with observing time at Keck Observatory and on the Automated Planet Finder.  We thank NASA for funding associated with our Key Strategic Mission Support project.  We gratefully acknowledge the efforts and dedication of the Keck Observatory staff for support of HIRES and remote observing.  We recognize and acknowledge the cultural role and reverence that the summit of Maunakea has within the indigenous Hawaiian community. We are deeply grateful to have the opportunity to conduct observations from this mountain.  We thank Ken and Gloria Levy, who supported the construction of the Levy Spectrometer on the Automated Planet Finder. We thank the University of California and Google for supporting Lick Observatory and the UCO staff for their dedicated work scheduling and operating the telescopes of Lick Observatory.  This paper is based on data collected by the TESS mission. Funding for the TESS mission is provided by the NASA Explorer Program. This research has made use of the Exoplanet Follow-up Observation Program (ExoFOP; DOI: 10.26134/ExoFOP5) website, which is operated by the California Institute of Technology, under contract with the National Aeronautics and Space Administration under the Exoplanet Exploration Program. This research has made use of the SIMBAD database, operated at CDS, Strasbourg, France.

Based on observations obtained at the international Gemini Observatory, a program of NSF’s NOIRLab, which is managed by the Association of Universities for Research in Astronomy (AURA) under a cooperative agreement with the National Science Foundation on behalf of the Gemini Observatory partnership: the National Science Foundation (United States), National Research Council (Canada), Agencia Nacional de Investigación y Desarrollo (Chile), Ministerio de Ciencia, Tecnología e Innovación (Argentina), Ministério da Ciência, Tecnologia, Inovações e Comunicações (Brazil), and Korea Astronomy and Space Science Institute (Republic of Korea).

This work has made use of data from the European Space Agency (ESA) mission
{\it Gaia} (\url{https://www.cosmos.esa.int/gaia}), processed by the {\it Gaia}
Data Processing and Analysis Consortium (DPAC,
\url{https://www.cosmos.esa.int/web/gaia/dpac/consortium}). Funding for the DPAC
has been provided by national institutions, in particular the institutions
participating in the {\it Gaia} Multilateral Agreement.

This paper includes data collected by the TESS mission that are publicly available from the Mikulski Archive for Space Telescopes (MAST). Data was taken from the TESS Input Catalog (doi:10.17909/fwdt-2x66).



\begin{thebibliography}{}
\expandafter\ifx\csname natexlab\endcsname\relax\def\natexlab#1{#1}\fi

\bibitem[{{Akeson} {et~al.}(2013){Akeson}, {Chen}, {Ciardi}, {Crane}, {Good},
  {Harbut}, {Jackson}, {Kane}, {Laity}, {Leifer}, {Lynn}, {McElroy}, {Papin},
  {Plavchan}, {Ram{\'\i}rez}, {Rey}, {von Braun}, {Wittman}, {Abajian}, {Ali},
  {Beichman}, {Beekley}, {Berriman}, {Berukoff}, {Bryden}, {Chan}, {Groom},
  {Lau}, {Payne}, {Regelson}, {Saucedo}, {Schmitz}, {Stauffer}, {Wyatt}, \&
  {Zhang}}]{akeson2013}
{Akeson}, R.~L., {Chen}, X., {Ciardi}, D., {et~al.} 2013, \pasp, 125, 989

\bibitem[{{Alibert} {et~al.}(2005){Alibert}, {Mordasini}, {Benz}, \&
  {Winisdoerffer}}]{alibert2005}
{Alibert}, Y., {Mordasini}, C., {Benz}, W., \& {Winisdoerffer}, C. 2005, \aap,
  434, 343

\bibitem[{{Barnes} {et~al.}(2015){Barnes}, {Deitrick}, {Greenberg}, {Quinn}, \&
  {Raymond}}]{Barnes2015}
{Barnes}, R., {Deitrick}, R., {Greenberg}, R., {Quinn}, T.~R., \& {Raymond},
  S.~N. 2015, \apj, 801, 101

\bibitem[{Bennett {et~al.}(2021)Bennett, Ranc, \& Fernandes}]{Bennett2021}
Bennett, D.~P., Ranc, C., \& Fernandes, R.~B. 2021, The Astronomical Journal,
  162, 243

\bibitem[{{Berger} {et~al.}(2020){Berger}, {Huber}, {van Saders}, {Gaidos},
  {Tayar}, \& {Kraus}}]{Berger2020}
{Berger}, T.~A., {Huber}, D., {van Saders}, J.~L., {et~al.} 2020, \aj, 159, 280

\bibitem[{{Bonomo} {et~al.}(2017){Bonomo}, {Desidera}, {Benatti}, {Borsa},
  {Crespi}, {Damasso}, {Lanza}, {Sozzetti}, {Lodato}, {Marzari}, {Boccato},
  {Claudi}, {Cosentino}, {Covino}, {Gratton}, {Maggio}, {Micela}, {Molinari},
  {Pagano}, {Piotto}, {Poretti}, {Smareglia}, {Affer}, {Biazzo}, {Bignamini},
  {Esposito}, {Giacobbe}, {H{\'e}brard}, {Malavolta}, {Maldonado}, {Mancini},
  {Martinez Fiorenzano}, {Masiero}, {Nascimbeni}, {Pedani}, {Rainer}, \& {Scand
  ariato}}]{bonomo2017}
{Bonomo}, A.~S., {Desidera}, S., {Benatti}, S., {et~al.} 2017, \aap, 602, A107

\bibitem[{{Borucki} {et~al.}(2010){Borucki}, {Koch}, {Basri}, {Batalha},
  {Brown}, {Caldwell}, {Caldwell}, {Christensen-Dalsgaard}, {Cochran},
  {DeVore}, {Dunham}, {Dupree}, {Gautier}, {Geary}, {Gilliland}, {Gould},
  {Howell}, {Jenkins}, {Kondo}, {Latham}, {Marcy}, {Meibom}, {Kjeldsen},
  {Lissauer}, {Monet}, {Morrison}, {Sasselov}, {Tarter}, {Boss}, {Brownlee},
  {Owen}, {Buzasi}, {Charbonneau}, {Doyle}, {Fortney}, {Ford}, {Holman},
  {Seager}, {Steffen}, {Welsh}, {Rowe}, {Anderson}, {Buchhave}, {Ciardi},
  {Walkowicz}, {Sherry}, {Horch}, {Isaacson}, {Everett}, {Fischer}, {Torres},
  {Johnson}, {Endl}, {MacQueen}, {Bryson}, {Dotson}, {Haas}, {Kolodziejczak},
  {Van Cleve}, {Chandrasekaran}, {Twicken}, {Quintana}, {Clarke}, {Allen},
  {Li}, {Wu}, {Tenenbaum}, {Verner}, {Bruhweiler}, {Barnes}, \&
  {Prsa}}]{borucki2010a}
{Borucki}, W.~J., {Koch}, D., {Basri}, G., {et~al.} 2010, Science, 327, 977

\bibitem[{Buchhave {et~al.}(2012)Buchhave, Latham, Johansen, Bizzarro, Torres,
  Rowe, Batalha, Borucki, Brugamyer, Caldwell, Bryson, Ciardi, Cochran, Endl,
  Esquerdo, Ford, Geary, Gilliland, Hansen, Isaacson, Laird, Lucas, Marcy,
  Morse, Robertson, Shporer, Stefanik, Still, \& Quinn}]{buchhave2012}
Buchhave, L.~A., Latham, D.~W., Johansen, A., {et~al.} 2012, Nature, 486, 375

\bibitem[{{Butler} {et~al.}(1996){Butler}, {Marcy}, {Williams}, {McCarthy},
  {Dosanjh}, \& {Vogt}}]{butler1996}
{Butler}, R.~P., {Marcy}, G.~W., {Williams}, E., {et~al.} 1996, \pasp, 108, 500

\bibitem[{{Carrera} {et~al.}(2019){Carrera}, {Raymond}, \&
  {Davies}}]{carrera2019b}
{Carrera}, D., {Raymond}, S.~N., \& {Davies}, M.~B. 2019, \aap, 629, L7

\bibitem[{{Chambers} {et~al.}(1996){Chambers}, {Wetherill}, \&
  {Boss}}]{chambers1996}
{Chambers}, J.~E., {Wetherill}, G.~W., \& {Boss}, A.~P. 1996, \icarus, 119, 261

\bibitem[{{Chatterjee} {et~al.}(2008){Chatterjee}, {Ford}, {Matsumura}, \&
  {Rasio}}]{chatterjee2008}
{Chatterjee}, S., {Ford}, E.~B., {Matsumura}, S., \& {Rasio}, F.~A. 2008, \apj,
  686, 580

\bibitem[{{Choi} {et~al.}(2016){Choi}, {Dotter}, {Conroy}, {Cantiello},
  {Paxton}, \& {Johnson}}]{Choi2016}
{Choi}, J., {Dotter}, A., {Conroy}, C., {et~al.} 2016, \apj, 823, 102

\bibitem[{Chontos {et~al.}(2022)Chontos, Murphy, MacDougall, Fetherolf, Zandt,
  Rubenzahl, Beard, Huber, Batalha, Crossfield, Dressing, Fulton, Howard,
  Isaacson, Kane, Petigura, Robertson, Roy, Weiss, Behmard, Dai, Dalba,
  Giacalone, Hill, Lubin, Mayo, Mo{\v{c}}nik, Polanski, Rosenthal, Scarsdale,
  Turtelboom, Ricker, Vanderspek, Latham, Seager, Winn, Jenkins, Quinn,
  Guerrero, Collins, Ciardi, Shporer, Goeke, Levine, Ting, Bieryla, Collins,
  Kielkopf, Barkaoui, Benni, Esparza-Borges, Conti, Hooton, Kagetani, Laloum,
  Marino, Massey, Murgas, Papini, Schwarz, Srdoc, Stockdale, Wang, Wittrock, \&
  Zou}]{Chontos2022}
Chontos, A., Murphy, J. M.~A., MacDougall, M.~G., {et~al.} 2022, The
  Astronomical Journal, 163, 297

\bibitem[{{Ciardi} {et~al.}(2015){Ciardi}, {Beichman}, {Horch}, \&
  {Howell}}]{Ciardi2015}
{Ciardi}, D.~R., {Beichman}, C.~A., {Horch}, E.~P., \& {Howell}, S.~B. 2015,
  \apj, 805, 16

\bibitem[{{Cloutier} {et~al.}(2018){Cloutier}, {Doyon}, {Bouchy}, \&
  {H{\'e}brard}}]{cloutier2018}
{Cloutier}, R., {Doyon}, R., {Bouchy}, F., \& {H{\'e}brard}, G. 2018, \aj, 156,
  82

\bibitem[{{Dai} {et~al.}(2020){Dai}, {Roy}, {Fulton}, {Robertson}, {Hirsch},
  {Isaacson}, {Albrecht}, {Mann}, {Kristiansen}, {Batalha}, {Beard}, {Behmard},
  {Chontos}, {Crossfield}, {Dalba}, {Dressing}, {Giacalone}, {Hill}, {Howard},
  {Huber}, {Kane}, {Kosiarek}, {Lubin}, {Mayo}, {Mocnik}, {Akana Murphy},
  {Petigura}, {Rosenthal}, {Rubenzahl}, {Scarsdale}, {Weiss}, {Van Zandt},
  {Ricker}, {Vanderspek}, {Latham}, {Seager}, {Winn}, {Jenkins}, {Caldwell},
  {Charbonneau}, {Daylan}, {G{\"u}nther}, {Morgan}, {Quinn}, {Rose}, \&
  {Smith}}]{Dai2020}
{Dai}, F., {Roy}, A., {Fulton}, B., {et~al.} 2020, \aj, 160, 193

\bibitem[{{Dai} {et~al.}(2021){Dai}, {Howard}, {Batalha}, {Beard}, {Behmard},
  {Blunt}, {Brinkman}, {Chontos}, {Crossfield}, {Dalba}, {Dressing}, {Fulton},
  {Giacalone}, {Hill}, {Huber}, {Isaacson}, {Kane}, {Lubin}, {Mayo},
  {Mo{\v{c}}nik}, {Akana Murphy}, {Petigura}, {Rice}, {Robertson}, {Rosenthal},
  {Roy}, {Rubenzahl}, {Weiss}, {Zandt}, {Beichman}, {Ciardi}, {Collins},
  {Gonzales}, {Howell}, {Matson}, {Matthews}, {Schlieder}, {Schwarz}, {Ricker},
  {Vanderspek}, {Latham}, {Seager}, {Winn}, {Jenkins}, {Caldwell}, {Colon},
  {Dragomir}, {Lund}, {McLean}, {Rudat}, \& {Shporer}}]{Dai2021}
{Dai}, F., {Howard}, A.~W., {Batalha}, N.~M., {et~al.} 2021, \aj, 162, 62

\bibitem[{{Dalba} {et~al.}(2020{\natexlab{a}}){Dalba}, {Fulton}, {Isaacson},
  {Kane}, \& {Howard}}]{dalba2020b}
{Dalba}, P.~A., {Fulton}, B., {Isaacson}, H., {Kane}, S.~R., \& {Howard}, A.~W.
  2020{\natexlab{a}}, \aj, 160, 149

\bibitem[{{Dalba} {et~al.}(2019){Dalba}, {Kane}, {Barclay}, {Bean}, {Campante},
  {Pepper}, {Ragozzine}, \& {Turnbull}}]{dalba2019c}
{Dalba}, P.~A., {Kane}, S.~R., {Barclay}, T., {et~al.} 2019, \pasp, 131, 034401

\bibitem[{{Dalba} {et~al.}(2020{\natexlab{b}}){Dalba}, {Gupta}, {Rodriguez},
  {Dragomir}, {Huang}, {Kane}, {Quinn}, {Bieryla}, {Esquerdo}, {Fulton},
  {Scarsdale}, {Batalha}, {Beard}, {Behmard}, {Chontos}, {Crossfield},
  {Dressing}, {Giacalone}, {Hill}, {Hirsch}, {Howard}, {Huber}, {Isaacson},
  {Kosiarek}, {Lubin}, {Mayo}, {Mocnik}, {Akana Murphy}, {Petigura},
  {Robertson}, {Rosenthal}, {Roy}, {Rubenzahl}, {Van Zandt}, {Weiss},
  {Knudstrup}, {Andersen}, {Grundahl}, {Yao}, {Pepper}, {Villanueva}, {Ciardi},
  {Cloutier}, {Jacobs}, {Kristiansen}, {LaCourse}, {Lendl}, {Osborn}, {Palle},
  {Stassun}, {Stevens}, {Ricker}, {Vanderspek}, {Latham}, {Seager}, {Winn},
  {Jenkins}, {Caldwell}, {Daylan}, {Fong}, {Goeke}, {Rose}, {Rowden},
  {Schlieder}, {Smith}, \& {Vanderburg}}]{dalba2020a}
{Dalba}, P.~A., {Gupta}, A.~F., {Rodriguez}, J.~E., {et~al.}
  2020{\natexlab{b}}, \aj, 159, 241

\bibitem[{Dalba {et~al.}(2022)Dalba, Kane, Dragomir, Villanueva, Collins,
  Jacobs, LaCourse, Gagliano, Kristiansen, Omohundro, Schwengeler, Terentev,
  Vanderburg, Fulton, Isaacson, Zandt, Howard, Thorngren, Howell, Batalha,
  Chontos, Crossfield, Dressing, Huber, Petigura, Robertson, Roy, Weiss,
  Behmard, Beard, Brinkman, Giacalone, Hill, Lubin, Mayo, Mo{\v{c}}nik, Murphy,
  Polanski, Rice, Rosenthal, Rubenzahl, Scarsdale, Turtelboom, Tyler, Benni,
  Boyce, Esposito, Girardin, Laloum, Lewin, Mann, Marchis, Schwarz, Srdoc,
  Steuer, Sivarani, Unni, Eisner, Fetherolf, Li, Yao, Pepper, Ricker,
  Vanderspek, Latham, Seager, Winn, Jenkins, Burke, Eastman, Lund, Rodriguez,
  Rowden, Ting, \& Villase{\~{n}}or}]{Dalba2022}
Dalba, P.~A., Kane, S.~R., Dragomir, D., {et~al.} 2022, The Astronomical
  Journal, 163, 61

\bibitem[{{Dragomir} {et~al.}(2020){Dragomir}, {Harris}, {Pepper}, {Barclay},
  {Villanueva}, {Ricker}, {Vanderspek}, {Latham}, {Seager}, {Winn}, {Jenkins},
  {Ciardi}, {Furesz}, {Henze}, {Mireles}, {Morgan}, {Quintana}, {Ting}, \&
  {Yahalomi}}]{dragomir2020}
{Dragomir}, D., {Harris}, M., {Pepper}, J., {et~al.} 2020, \aj, 159, 219

\bibitem[{{Dulz} {et~al.}(2020){Dulz}, {Plavchan}, {Crepp}, {Stark}, {Morgan},
  {Kane}, {Newman}, {Matzko}, \& {Mulders}}]{dulz2020}
{Dulz}, S.~D., {Plavchan}, P., {Crepp}, J.~R., {et~al.} 2020, \apj, 893, 122

\bibitem[{{Duncan} {et~al.}(1998){Duncan}, {Levison}, \& {Lee}}]{duncan1998}
{Duncan}, M.~J., {Levison}, H.~F., \& {Lee}, M.~H. 1998, \aj, 116, 2067

\bibitem[{{Eastman} {et~al.}(2013){Eastman}, {Gaudi}, \& {Agol}}]{eastman2013}
{Eastman}, J., {Gaudi}, B.~S., \& {Agol}, E. 2013, \pasp, 125, 83

\bibitem[{{Eastman} {et~al.}(2019){Eastman}, {Rodriguez}, {Agol}, {Stassun},
  {Beatty}, {Vanderburg}, {Gaudi}, {Collins}, \& {Luger}}]{eastman2020}
{Eastman}, J.~D., {Rodriguez}, J.~E., {Agol}, E., {et~al.} 2019, arXiv
  e-prints, arXiv:1907.09480

\bibitem[{{Emsenhuber} {et~al.}(2021){Emsenhuber}, {Mordasini}, {Burn},
  {Alibert}, {Benz}, \& {Asphaug}}]{Emsenhuber2021}
{Emsenhuber}, A., {Mordasini}, C., {Burn}, R., {et~al.} 2021, \aap, 656, A69

\bibitem[{{Ford}(2014)}]{ford2014}
{Ford}, E.~B. 2014, Proceedings of the National Academy of Science, 111, 12616

\bibitem[{{Foreman-Mackey} {et~al.}(2013){Foreman-Mackey}, {Hogg}, {Lang}, \&
  {Goodman}}]{Foreman2013}
{Foreman-Mackey}, D., {Hogg}, D.~W., {Lang}, D., \& {Goodman}, J. 2013, \pasp,
  125, 306

\bibitem[{{Fulton} {et~al.}(2018){Fulton}, {Petigura}, {Blunt}, \&
  {Sinukoff}}]{fulton2018a}
{Fulton}, B.~J., {Petigura}, E.~A., {Blunt}, S., \& {Sinukoff}, E. 2018, \pasp,
  130, 044504

\bibitem[{{Fulton} {et~al.}(2017){Fulton}, {Petigura}, {Howard}, {Isaacson},
  {Marcy}, {Cargile}, {Hebb}, {Weiss}, {Johnson}, {Morton}, {Sinukoff},
  {Crossfield}, \& {Hirsch}}]{fulton2017}
{Fulton}, B.~J., {Petigura}, E.~A., {Howard}, A.~W., {et~al.} 2017, \aj, 154,
  109
  
 \bibitem[{{Furesz}(2008)}]{Furesz2008}
{Furesz}, G. 2008, PhD thesis University of Szeged

\bibitem[{{Gaia Collaboration} {et~al.}(2018){Gaia Collaboration}, {Brown},
  {Vallenari}, {Prusti}, {de Bruijne}, {Babusiaux}, {Bailer-Jones}, {Biermann},
  {Evans}, {Eyer}, {Jansen}, {Jordi}, {Klioner}, {Lammers}, {Lindegren},
  {Luri}, {Mignard}, {Panem}, {Pourbaix}, {Randich}, {Sartoretti}, {Siddiqui},
  {Soubiran}, {van Leeuwen}, {Walton}, {Arenou}, {Bastian}, {Cropper},
  {Drimmel}, {Katz}, {Lattanzi}, {Bakker}, {Cacciari}, {Casta{\~n}eda},
  {Chaoul}, {Cheek}, {De Angeli}, {Fabricius}, {Guerra}, {Holl}, {Masana},
  {Messineo}, {Mowlavi}, {Nienartowicz}, {Panuzzo}, {Portell}, {Riello},
  {Seabroke}, {Tanga}, {Th{\'e}venin}, {Gracia-Abril}, {Comoretto},
  {Garcia-Reinaldos}, {Teyssier}, {Altmann}, {Andrae}, {Audard},
  {Bellas-Velidis}, {Benson}, {Berthier}, {Blomme}, {Burgess}, {Busso},
  {Carry}, {Cellino}, {Clementini}, {Clotet}, {Creevey}, {Davidson}, {De
  Ridder}, {Delchambre}, {Dell'Oro}, {Ducourant},
  {Fern{\'a}ndez-Hern{\'a}ndez}, {Fouesneau}, {Fr{\'e}mat}, {Galluccio},
  {Garc{\'\i}a-Torres}, {Gonz{\'a}lez-N{\'u}{\~n}ez}, {Gonz{\'a}lez-Vidal},
  {Gosset}, {Guy}, {Halbwachs}, {Hambly}, {Harrison}, {Hern{\'a}ndez},
  {Hestroffer}, {Hodgkin}, {Hutton}, {Jasniewicz}, {Jean-Antoine-Piccolo},
  {Jordan}, {Korn}, {Krone-Martins}, {Lanzafame}, {Lebzelter}, {L{\"o}ffler},
  {Manteiga}, {Marrese}, {Mart{\'\i}n-Fleitas}, {Moitinho}, {Mora}, {Muinonen},
  {Osinde}, {Pancino}, {Pauwels}, {Petit}, {Recio-Blanco}, {Richards},
  {Rimoldini}, {Robin}, {Sarro}, {Siopis}, {Smith}, {Sozzetti}, {S{\"u}veges},
  {Torra}, {van Reeven}, {Abbas}, {Abreu Aramburu}, {Accart}, {Aerts},
  {Altavilla}, {{\'A}lvarez}, {Alvarez}, {Alves}, {Anderson}, {Andrei},
  {Anglada Varela}, {Antiche}, {Antoja}, {Arcay}, {Astraatmadja}, {Bach},
  {Baker}, {Balaguer-N{\'u}{\~n}ez}, {Balm}, {Barache}, {Barata}, {Barbato},
  {Barblan}, {Barklem}, {Barrado}, {Barros}, {Barstow}, {Bartholom{\'e}
  Mu{\~n}oz}, {Bassilana}, {Becciani}, {Bellazzini}, {Berihuete}, {Bertone},
  {Bianchi}, {Bienaym{\'e}}, {Blanco-Cuaresma}, {Boch}, {Boeche}, {Bombrun},
  {Borrachero}, {Bossini}, {Bouquillon}, {Bourda}, {Bragaglia}, {Bramante},
  {Breddels}, {Bressan}, {Brouillet}, {Br{\"u}semeister}, {Brugaletta},
  {Bucciarelli}, {Burlacu}, {Busonero}, {Butkevich}, {Buzzi}, {Caffau},
  {Cancelliere}, {Cannizzaro}, {Cantat-Gaudin}, {Carballo}, {Carlucci},
  {Carrasco}, {Casamiquela}, {Castellani}, {Castro-Ginard}, {Charlot},
  {Chemin}, {Chiavassa}, {Cocozza}, {Costigan}, {Cowell}, {Crifo}, {Crosta},
  {Crowley}, {Cuypers}, {Dafonte}, {Damerdji}, {Dapergolas}, {David}, {David},
  {de Laverny}, {De Luise}, {De March}, {de Martino}, {de Souza}, {de Torres},
  {Debosscher}, {del Pozo}, {Delbo}, {Delgado}, {Delgado}, {Di Matteo},
  {Diakite}, {Diener}, {Distefano}, {Dolding}, {Drazinos}, {Dur{\'a}n},
  {Edvardsson}, {Enke}, {Eriksson}, {Esquej}, {Eynard Bontemps}, {Fabre},
  {Fabrizio}, {Faigler}, {Falc{\~a}o}, {Farr{\`a}s Casas}, {Federici},
  {Fedorets}, {Fernique}, {Figueras}, {Filippi}, {Findeisen}, {Fonti},
  {Fraile}, {Fraser}, {Fr{\'e}zouls}, {Gai}, {Galleti}, {Garabato},
  {Garc{\'\i}a-Sedano}, {Garofalo}, {Garralda}, {Gavel}, {Gavras}, {Gerssen},
  {Geyer}, {Giacobbe}, {Gilmore}, {Girona}, {Giuffrida}, {Glass}, {Gomes},
  {Granvik}, {Gueguen}, {Guerrier}, {Guiraud}, {Guti{\'e}rrez-S{\'a}nchez},
  {Haigron}, {Hatzidimitriou}, {Hauser}, {Haywood}, {Heiter}, {Helmi}, {Heu},
  {Hilger}, {Hobbs}, {Hofmann}, {Holland}, {Huckle}, {Hypki}, {Icardi},
  {Jan{\ss}en}, {Jevardat de Fombelle}, {Jonker}, {Juh{\'a}sz}, {Julbe},
  {Karampelas}, {Kewley}, {Klar}, {Kochoska}, {Kohley}, {Kolenberg},
  {Kontizas}, {Kontizas}, {Koposov}, {Kordopatis}, {Kostrzewa-Rutkowska},
  {Koubsky}, {Lambert}, {Lanza}, {Lasne}, {Lavigne}, {Le Fustec}, {Le
  Poncin-Lafitte}, {Lebreton}, {Leccia}, {Leclerc}, {Lecoeur-Taibi},
  {Lenhardt}, {Leroux}, {Liao}, {Licata}, {Lindstr{\o}m}, {Lister}, {Livanou},
  {Lobel}, {L{\'o}pez}, {Managau}, {Mann}, {Mantelet}, {Marchal}, {Marchant},
  {Marconi}, {Marinoni}, {Marschalk{\'o}}, {Marshall}, {Martino}, {Marton},
  {Mary}, {Massari}, {Matijevi{\v{c}}}, {Mazeh}, {McMillan}, {Messina},
  {Michalik}, {Millar}, {Molina}, {Molinaro}, {Moln{\'a}r}, {Montegriffo},
  {Mor}, {Morbidelli}, {Morel}, {Morris}, {Mulone}, {Muraveva}, {Musella},
  {Nelemans}, {Nicastro}, {Noval}, {O'Mullane}, {Ord{\'e}novic},
  {Ord{\'o}{\~n}ez-Blanco}, {Osborne}, {Pagani}, {Pagano}, {Pailler},
  {Palacin}, {Palaversa}, {Panahi}, {Pawlak}, {Piersimoni}, {Pineau}, {Plachy},
  {Plum}, {Poggio}, {Poujoulet}, {Pr{\v{s}}a}, {Pulone}, {Racero}, {Ragaini},
  {Rambaux}, {Ramos-Lerate}, {Regibo}, {Reyl{\'e}}, {Riclet}, {Ripepi}, {Riva},
  {Rivard}, {Rixon}, {Roegiers}, {Roelens}, {Romero-G{\'o}mez}, {Rowell},
  {Royer}, {Ruiz-Dern}, {Sadowski}, {Sagrist{\`a} Sell{\'e}s}, {Sahlmann},
  {Salgado}, {Salguero}, {Sanna}, {Santana-Ros}, {Sarasso}, {Savietto},
  {Schultheis}, {Sciacca}, {Segol}, {Segovia}, {S{\'e}gransan}, {Shih},
  {Siltala}, {Silva}, {Smart}, {Smith}, {Solano}, {Solitro}, {Sordo}, {Soria
  Nieto}, {Souchay}, {Spagna}, {Spoto}, {Stampa}, {Steele},
  {Steidelm{\"u}ller}, {Stephenson}, {Stoev}, {Suess}, {Surdej}, {Szabados},
  {Szegedi-Elek}, {Tapiador}, {Taris}, {Tauran}, {Taylor}, {Teixeira},
  {Terrett}, {Teyssand ier}, {Thuillot}, {Titarenko}, {Torra Clotet}, {Turon},
  {Ulla}, {Utrilla}, {Uzzi}, {Vaillant}, {Valentini}, {Valette}, {van Elteren},
  {Van Hemelryck}, {van Leeuwen}, {Vaschetto}, {Vecchiato}, {Veljanoski},
  {Viala}, {Vicente}, {Vogt}, {von Essen}, {Voss}, {Votruba}, {Voutsinas},
  {Walmsley}, {Weiler}, {Wertz}, {Wevers}, {Wyrzykowski}, {Yoldas},
  {{\v{Z}}erjal}, {Ziaeepour}, {Zorec}, {Zschocke}, {Zucker}, {Zurbach}, \&
  {Zwitter}}]{GaiaCollab}
{Gaia Collaboration}, {Brown}, A.~G.~A., {Vallenari}, A., {et~al.} 2018, \aap,
  616, A1

\bibitem[{{Gaia Collaboration} {et~al.}(2021){Gaia Collaboration}, {Brown},
  {Vallenari}, {Prusti}, {de Bruijne}, {Babusiaux}, {Biermann}, {Creevey},
  {Evans}, {Eyer}, {Hutton}, {Jansen}, {Jordi}, {Klioner}, {Lammers},
  {Lindegren}, {Luri}, {Mignard}, {Panem}, {Pourbaix}, {Randich}, {Sartoretti},
  {Soubiran}, {Walton}, {Arenou}, {Bailer-Jones}, {Bastian}, {Cropper},
  {Drimmel}, {Katz}, {Lattanzi}, {van Leeuwen}, {Bakker}, {Cacciari},
  {Casta{\~n}eda}, {De Angeli}, {Ducourant}, {Fabricius}, {Fouesneau},
  {Fr{\'e}mat}, {Guerra}, {Guerrier}, {Guiraud}, {Jean-Antoine Piccolo},
  {Masana}, {Messineo}, {Mowlavi}, {Nicolas}, {Nienartowicz}, {Pailler},
  {Panuzzo}, {Riclet}, {Roux}, {Seabroke}, {Sordo}, {Tanga}, {Th{\'e}venin},
  {Gracia-Abril}, {Portell}, {Teyssier}, {Altmann}, {Andrae}, {Bellas-Velidis},
  {Benson}, {Berthier}, {Blomme}, {Brugaletta}, {Burgess}, {Busso}, {Carry},
  {Cellino}, {Cheek}, {Clementini}, {Damerdji}, {Davidson}, {Delchambre},
  {Dell'Oro}, {Fern{\'a}ndez-Hern{\'a}ndez}, {Galluccio}, {Garc{\'\i}a-Lario},
  {Garcia-Reinaldos}, {Gonz{\'a}lez-N{\'u}{\~n}ez}, {Gosset}, {Haigron},
  {Halbwachs}, {Hambly}, {Harrison}, {Hatzidimitriou}, {Heiter},
  {Hern{\'a}ndez}, {Hestroffer}, {Hodgkin}, {Holl}, {Jan{\ss}en}, {Jevardat de
  Fombelle}, {Jordan}, {Krone-Martins}, {Lanzafame}, {L{\"o}ffler}, {Lorca},
  {Manteiga}, {Marchal}, {Marrese}, {Moitinho}, {Mora}, {Muinonen}, {Osborne},
  {Pancino}, {Pauwels}, {Petit}, {Recio-Blanco}, {Richards}, {Riello},
  {Rimoldini}, {Robin}, {Roegiers}, {Rybizki}, {Sarro}, {Siopis}, {Smith},
  {Sozzetti}, {Ulla}, {Utrilla}, {van Leeuwen}, {van Reeven}, {Abbas}, {Abreu
  Aramburu}, {Accart}, {Aerts}, {Aguado}, {Ajaj}, {Altavilla}, {{\'A}lvarez},
  {{\'A}lvarez Cid-Fuentes}, {Alves}, {Anderson}, {Anglada Varela}, {Antoja},
  {Audard}, {Baines}, {Baker}, {Balaguer-N{\'u}{\~n}ez}, {Balbinot}, {Balog},
  {Barache}, {Barbato}, {Barros}, {Barstow}, {Bartolom{\'e}}, {Bassilana},
  {Bauchet}, {Baudesson-Stella}, {Becciani}, {Bellazzini}, {Bernet}, {Bertone},
  {Bianchi}, {Blanco-Cuaresma}, {Boch}, {Bombrun}, {Bossini}, {Bouquillon},
  {Bragaglia}, {Bramante}, {Breedt}, {Bressan}, {Brouillet}, {Bucciarelli},
  {Burlacu}, {Busonero}, {Butkevich}, {Buzzi}, {Caffau}, {Cancelliere},
  {C{\'a}novas}, {Cantat-Gaudin}, {Carballo}, {Carlucci}, {Carnerero},
  {Carrasco}, {Casamiquela}, {Castellani}, {Castro-Ginard}, {Castro Sampol},
  {Chaoul}, {Charlot}, {Chemin}, {Chiavassa}, {Cioni}, {Comoretto}, {Cooper},
  {Cornez}, {Cowell}, {Crifo}, {Crosta}, {Crowley}, {Dafonte}, {Dapergolas},
  {David}, {David}, {de Laverny}, {De Luise}, {De March}, {De Ridder}, {de
  Souza}, {de Teodoro}, {de Torres}, {del Peloso}, {del Pozo}, {Delbo},
  {Delgado}, {Delgado}, {Delisle}, {Di Matteo}, {Diakite}, {Diener},
  {Distefano}, {Dolding}, {Eappachen}, {Edvardsson}, {Enke}, {Esquej}, {Fabre},
  {Fabrizio}, {Faigler}, {Fedorets}, {Fernique}, {Fienga}, {Figueras},
  {Fouron}, {Fragkoudi}, {Fraile}, {Franke}, {Gai}, {Garabato},
  {Garcia-Gutierrez}, {Garc{\'\i}a-Torres}, {Garofalo}, {Gavras}, {Gerlach},
  {Geyer}, {Giacobbe}, {Gilmore}, {Girona}, {Giuffrida}, {Gomel}, {Gomez},
  {Gonzalez-Santamaria}, {Gonz{\'a}lez-Vidal}, {Granvik},
  {Guti{\'e}rrez-S{\'a}nchez}, {Guy}, {Hauser}, {Haywood}, {Helmi}, {Hidalgo},
  {Hilger}, {H{\l}adczuk}, {Hobbs}, {Holland}, {Huckle}, {Jasniewicz},
  {Jonker}, {Juaristi Campillo}, {Julbe}, {Karbevska}, {Kervella}, {Khanna},
  {Kochoska}, {Kontizas}, {Kordopatis}, {Korn}, {Kostrzewa-Rutkowska},
  {Kruszy{\'n}ska}, {Lambert}, {Lanza}, {Lasne}, {Le Campion}, {Le Fustec},
  {Lebreton}, {Lebzelter}, {Leccia}, {Leclerc}, {Lecoeur-Taibi}, {Liao},
  {Licata}, {Lindstr{\o}m}, {Lister}, {Livanou}, {Lobel}, {Madrero Pardo},
  {Managau}, {Mann}, {Marchant}, {Marconi}, {Marcos Santos}, {Marinoni},
  {Marocco}, {Marshall}, {Martin Polo}, {Mart{\'\i}n-Fleitas}, {Masip},
  {Massari}, {Mastrobuono-Battisti}, {Mazeh}, {McMillan}, {Messina},
  {Michalik}, {Millar}, {Mints}, {Molina}, {Molinaro}, {Moln{\'a}r},
  {Montegriffo}, {Mor}, {Morbidelli}, {Morel}, {Morris}, {Mulone}, {Munoz},
  {Muraveva}, {Murphy}, {Musella}, {Noval}, {Ord{\'e}novic}, {Orr{\`u}},
  {Osinde}, {Pagani}, {Pagano}, {Palaversa}, {Palicio}, {Panahi}, {Pawlak},
  {Pe{\~n}alosa Esteller}, {Penttil{\"a}}, {Piersimoni}, {Pineau}, {Plachy},
  {Plum}, {Poggio}, {Poretti}, {Poujoulet}, {Pr{\v{s}}a}, {Pulone}, {Racero},
  {Ragaini}, {Rainer}, {Raiteri}, {Rambaux}, {Ramos}, {Ramos-Lerate}, {Re
  Fiorentin}, {Regibo}, {Reyl{\'e}}, {Ripepi}, {Riva}, {Rixon}, {Robichon},
  {Robin}, {Roelens}, {Rohrbasser}, {Romero-G{\'o}mez}, {Rowell}, {Royer},
  {Rybicki}, {Sadowski}, {Sagrist{\`a} Sell{\'e}s}, {Sahlmann}, {Salgado},
  {Salguero}, {Samaras}, {Sanchez Gimenez}, {Sanna}, {Santove{\~n}a},
  {Sarasso}, {Schultheis}, {Sciacca}, {Segol}, {Segovia}, {S{\'e}gransan},
  {Semeux}, {Shahaf}, {Siddiqui}, {Siebert}, {Siltala}, {Slezak}, {Smart},
  {Solano}, {Solitro}, {Souami}, {Souchay}, {Spagna}, {Spoto}, {Steele},
  {Steidelm{\"u}ller}, {Stephenson}, {S{\"u}veges}, {Szabados}, {Szegedi-Elek},
  {Taris}, {Tauran}, {Taylor}, {Teixeira}, {Thuillot}, {Tonello}, {Torra},
  {Torra}, {Turon}, {Unger}, {Vaillant}, {van Dillen}, {Vanel}, {Vecchiato},
  {Viala}, {Vicente}, {Voutsinas}, {Weiler}, {Wevers}, {Wyrzykowski}, {Yoldas},
  {Yvard}, {Zhao}, {Zorec}, {Zucker}, {Zurbach}, \& {Zwitter}}]{GaiaDR3_2021}
---. 2021, \aap, 650, C3

\bibitem[{{Gardner} {et~al.}(2006){Gardner}, {Mather}, {Clampin}, {Doyon},
  {Greenhouse}, {Hammel}, {Hutchings}, {Jakobsen}, {Lilly}, {Long}, {Lunine},
  {McCaughrean}, {Mountain}, {Nella}, {Rieke}, {Rieke}, {Rix}, {Smith},
  {Sonneborn}, {Stiavelli}, {Stockman}, {Windhorst}, \& {Wright}}]{gardner06}
{Gardner}, J.~P., {Mather}, J.~C., {Clampin}, M., {et~al.} 2006, \ssr, 123, 485

\bibitem[{{Guerrero} {et~al.}(2021){Guerrero}, {Seager}, {Huang}, {Vanderburg},
  {Garcia Soto}, {Mireles}, {Hesse}, {Fong}, {Glidden}, {Shporer}, {Latham},
  {Collins}, {Quinn}, {Burt}, {Dragomir}, {Crossfield}, {Vanderspek},
  {Fausnaugh}, {Burke}, {Ricker}, {Daylan}, {Essack}, {G{\"u}nther}, {Osborn},
  {Pepper}, {Rowden}, {Sha}, {Villanueva}, {Yahalomi}, {Yu}, {Ballard},
  {Batalha}, {Berardo}, {Chontos}, {Dittmann}, {Esquerdo}, {Mikal-Evans},
  {Jayaraman}, {Krishnamurthy}, {Louie}, {Mehrle}, {Niraula}, {Rackham},
  {Rodriguez}, {Rowden}, {Sousa-Silva}, {Watanabe}, {Wong}, {Zhan},
  {Zivanovic}, {Christiansen}, {Ciardi}, {Swain}, {Lund}, {Mullally},
  {Fleming}, {Rodriguez}, {Boyd}, {Quintana}, {Barclay}, {Col{\'o}n},
  {Rinehart}, {Schlieder}, {Clampin}, {Jenkins}, {Twicken}, {Caldwell},
  {Coughlin}, {Henze}, {Lissauer}, {Morris}, {Rose}, {Smith}, {Tenenbaum},
  {Ting}, {Wohler}, {Bakos}, {Bean}, {Berta-Thompson}, {Bieryla}, {Bouma},
  {Buchhave}, {Butler}, {Charbonneau}, {Doty}, {Ge}, {Holman}, {Howard},
  {Kaltenegger}, {Kane}, {Kjeldsen}, {Kreidberg}, {Lin}, {Minsky}, {Narita},
  {Paegert}, {P{\'a}l}, {Palle}, {Sasselov}, {Spencer}, {Sozzetti}, {Stassun},
  {Torres}, {Udry}, \& {Winn}}]{guerrero2021}
{Guerrero}, N.~M., {Seager}, S., {Huang}, C.~X., {et~al.} 2021, \apjs, 254, 39

\bibitem[{{Hastings}(1970)}]{hastings70}
{Hastings}, W.~K. 1970, Biometrika, 57, 97

\bibitem[{Hill {et~al.}(2023)Hill, Bott, Dalba, Fetherolf, Kane, Kopparapu, Li,
  \& Ostberg}]{Hill2023}
Hill, M.~L., Bott, K., Dalba, P.~A., {et~al.} 2023, The Astronomical Journal,
  165, 34

\bibitem[{{Hodapp} {et~al.}(2003){Hodapp}, {Jensen}, {Irwin}, {Yamada},
  {Chung}, {Fletcher}, {Robertson}, {Hora}, {Simons}, {Mays}, {Nolan}, {Bec},
  {Merrill}, \& {Fowler}}]{hodapp2003}
{Hodapp}, K.~W., {Jensen}, J.~B., {Irwin}, E.~M., {et~al.} 2003, \pasp, 115,
  1388

\bibitem[{{Horner} {et~al.}(2020){Horner}, {Kane}, {Marshall}, {Dalba}, {Holt},
  {Wood}, {Maynard-Casely}, {Wittenmyer}, {Lykawka}, {Hill}, {Salmeron},
  {Bailey}, {L{\"o}hne}, {Agnew}, {Carter}, \& {Tylor}}]{horner2020b}
{Horner}, J., {Kane}, S.~R., {Marshall}, J.~P., {et~al.} 2020, \pasp, 132,
  102001

\bibitem[{Howard {et~al.}(2010)Howard, Johnson, Marcy, Fischer, Wright, Bernat,
  Henry, Peek, Isaacson, Apps, Endl, Cochran, Valenti, Anderson, \&
  Piskunov}]{Howard_2010}
Howard, A.~W., Johnson, J.~A., Marcy, G.~W., {et~al.} 2010, The Astrophysical
  Journal, 721, 1467

\bibitem[{{Howell} {et~al.}(2011){Howell}, {Everett}, {Sherry}, {Horch}, \&
  {Ciardi}}]{Howell2011}
{Howell}, S.~B., {Everett}, M.~E., {Sherry}, W., {Horch}, E., \& {Ciardi},
  D.~R. 2011, \aj, 142, 19

\bibitem[{{Huang} {et~al.}(2020{\natexlab{a}}){Huang}, {Vanderburg}, {P{\'a}l},
  {Sha}, {Yu}, {Fong}, {Fausnaugh}, {Shporer}, {Guerrero}, {Vanderspek}, \&
  {Ricker}}]{huang2020a}
{Huang}, C.~X., {Vanderburg}, A., {P{\'a}l}, A., {et~al.} 2020{\natexlab{a}},
  Research Notes of the American Astronomical Society, 4, 204

\bibitem[{{Huang} {et~al.}(2020{\natexlab{b}}){Huang}, {Vanderburg}, {P{\'a}l},
  {Sha}, {Yu}, {Fong}, {Fausnaugh}, {Shporer}, {Guerrero}, {Vanderspek}, \&
  {Ricker}}]{huang2020b}
---. 2020{\natexlab{b}}, Research Notes of the American Astronomical Society,
  4, 206

\bibitem[{{Huber} {et~al.}(2017){Huber}, {Zinn}, {Bojsen-Hansen},
  {Pinsonneault}, {Sahlholdt}, {Serenelli}, {Silva Aguirre}, {Stassun},
  {Stello}, {Tayar}, {Bastien}, {Bedding}, {Buchhave}, {Chaplin}, {Davies},
  {Garc{\'\i}a}, {Latham}, {Mathur}, {Mosser}, \& {Sharma}}]{Huber2017}
{Huber}, D., {Zinn}, J., {Bojsen-Hansen}, M., {et~al.} 2017, \apj, 844, 102

\bibitem[{{Ida} \& {Lin}(2004)}]{Ida2004}
{Ida}, S., \& {Lin}, D.~N.~C. 2004, \apj, 604, 388

\bibitem[{{Isaacson} \& {Fischer}(2010)}]{IandF2010}
{Isaacson}, H., \& {Fischer}, D. 2010, \apj, 725, 875

\bibitem[{{Juri{\'c}} \& {Tremaine}(2008)}]{juric2008}
{Juri{\'c}}, M., \& {Tremaine}, S. 2008, \apj, 686, 603

\bibitem[{{Kane}(2016)}]{kane2016d}
{Kane}, S.~R. 2016, \apj, 830, 105

\bibitem[{{Kane} \& {Gelino}(2013)}]{kane2013b}
{Kane}, S.~R., \& {Gelino}, D.~M. 2013, \apj, 762, 129

\bibitem[{{Kane} {et~al.}(2021{\natexlab{a}}){Kane}, {Li}, {Wolf}, {Ostberg},
  \& {Hill}}]{kane2021a}
{Kane}, S.~R., {Li}, Z., {Wolf}, E.~T., {Ostberg}, C., \& {Hill}, M.~L.
  2021{\natexlab{a}}, \aj, 161, 31

\bibitem[{{Kane} {et~al.}(2009){Kane}, {Mahadevan}, {von Braun}, {Laughlin}, \&
  {Ciardi}}]{kane2009c}
{Kane}, S.~R., {Mahadevan}, S., {von Braun}, K., {Laughlin}, G., \& {Ciardi},
  D.~R. 2009, \pasp, 121, 1386

\bibitem[{{Kane} {et~al.}(2021{\natexlab{b}}){Kane}, {Bean}, {Campante}, {D
  alba}, {Fetherolf}, {Mocnik}, {Ostberg}, {Pepper}, {Simpson}, {Turnbull},
  {Ricker}, {Latham}, {Seager}, {Winn}, {Huber}, \& {Chaplin}}]{kane2021b}
{Kane}, S.~R., {Bean}, J.~L., {Campante}, T.~L., {et~al.} 2021{\natexlab{b}},
  \pasp, 133, 014402

\bibitem[{{Kane} {et~al.}(2021{\natexlab{c}}){Kane}, {Arney}, {Byrne}, {Dalba},
  {Desch}, {Horner}, {Izenberg}, {Mandt}, {Meadows}, \& {Quick}}]{kane2021d}
{Kane}, S.~R., {Arney}, G.~N., {Byrne}, P.~K., {et~al.} 2021{\natexlab{c}},
  Journal of Geophysical Research (Planets), 126, e06643

\bibitem[{{Kane} {et~al.}(2023){Kane}, {Hill}, {Dalba}, {Fetherolf}, {Henry},
  {Fajardo-Acosta}, {Gnilka}, {Howard}, {Howell}, \& {Isaacson}}]{kane2023}
{Kane}, S.~R., {Hill}, M.~L., {Dalba}, P.~A., {et~al.} 2023, \aj, 165, 252

\bibitem[{{Kempton} {et~al.}(2018){Kempton}, {Bean}, {Louie}, {Deming}, {Koll},
  {Mansfield}, {Christiansen}, {L{\'o}pez-Morales}, {Swain}, {Zellem},
  {Ballard}, {Barclay}, {Barstow}, {Batalha}, {Beatty}, {Berta-Thompson},
  {Birkby}, {Buchhave}, {Charbonneau}, {Cowan}, {Crossfield}, {de Val-Borro},
  {Doyon}, {Dragomir}, {Gaidos}, {Heng}, {Hu}, {Kane}, {Kreidberg}, {Mallonn},
  {Morley}, {Narita}, {Nascimbeni}, {Pall{\'e}}, {Quintana}, {Rauscher},
  {Seager}, {Shkolnik}, {Sing}, {Sozzetti}, {Stassun}, {Valenti}, \& {von
  Essen}}]{kempton2018}
{Kempton}, E. M.~R., {Bean}, J.~L., {Louie}, D.~R., {et~al.} 2018, \pasp, 130,
  114401

\bibitem[{{Laughlin} \& {Chambers}(2001)}]{laughlin2001}
{Laughlin}, G., \& {Chambers}, J.~E. 2001, \apjl, 551, L109

\bibitem[{{Lightkurve Collaboration} {et~al.}(2018){Lightkurve Collaboration},
  {Cardoso}, {Hedges}, {Gully-Santiago}, {Saunders}, {Cody}, {Barclay}, {Hall},
  {Sagear}, {Turtelboom}, {Zhang}, {Tzanidakis}, {Mighell}, {Coughlin}, {Bell},
  {Berta-Thompson}, {Williams}, {Dotson}, \& {Barentsen}}]{lightkurve}
{Lightkurve Collaboration}, {Cardoso}, J. V. d. M.~a., {Hedges}, C., {et~al.}
  2018, {Lightkurve: Kepler and TESS time series analysis in Python}, , ,
  ascl:1812.013

\bibitem[{Lubin {et~al.}(2022)Lubin, Zandt, Holcomb, Weiss, Petigura,
  Robertson, Murphy, Scarsdale, Batygin, Polanski, Batalha, Crossfield,
  Dressing, Fulton, Howard, Huber, Isaacson, Kane, Roy, Beard, Blunt, Chontos,
  Dai, Dalba, Gary, Giacalone, Hill, Mayo, Mo{\v{c}}nik, Kosiarek, Rice,
  Rubenzahl, Latham, Seager, Winn, \& Gary}]{Lubin2022}
Lubin, J., Zandt, J.~V., Holcomb, R., {et~al.} 2022, The Astronomical Journal,
  163, 101

\bibitem[{MacDougall {et~al.}(2021)MacDougall, Petigura, Angelo, Lubin,
  Batalha, Beard, Behmard, Blunt, Brinkman, Chontos, Crossfield, Dai, Dalba,
  Dressing, Fulton, Giacalone, Hill, Howard, Huber, Isaacson, Kane, Mayo,
  Mo{\v{c}}nik, Murphy, Polanski, Rice, Robertson, Rosenthal, Roy, Rubenzahl,
  Scarsdale, Turtelboom, Zandt, Weiss, Matthews, Jenkins, Latham, Ricker,
  Seager, Vanderspek, Winn, Brasseur, Doty, Fausnaugh, Guerrero, Henze, Lund,
  \& Shporer}]{MacDougall2021}
MacDougall, M.~G., Petigura, E.~A., Angelo, I., {et~al.} 2021, The Astronomical
  Journal, 162, 265

\bibitem[{{Mayor} {et~al.}(2011){Mayor}, {Marmier}, {Lovis}, {Udry},
  {S{\'e}gransan}, {Pepe}, {Benz}, {Bertaux}, {Bouchy}, {Dumusque}, {Lo Curto},
  {Mordasini}, {Queloz}, \& {Santos}}]{Mayor2011}
{Mayor}, M., {Marmier}, M., {Lovis}, C., {et~al.} 2011, arXiv e-prints,
  arXiv:1109.2497

\bibitem[{{Metropolis} {et~al.}(1953){Metropolis}, {Rosenbluth}, {Rosenbluth},
  {Teller}, \& {Teller}}]{metropolis53}
{Metropolis}, N., {Rosenbluth}, A.~W., {Rosenbluth}, M.~N., {Teller}, A.~H., \&
  {Teller}, E. 1953, \jcp, 21, 1087

\bibitem[{{Mishra} {et~al.}(2023{\natexlab{a}}){Mishra}, {Alibert}, {Udry}, \&
  {Mordasini}}]{mishra2023a}
{Mishra}, L., {Alibert}, Y., {Udry}, S., \& {Mordasini}, C. 2023{\natexlab{a}},
  \aap, 670, A68

\bibitem[{{Mishra} {et~al.}(2023{\natexlab{b}}){Mishra}, {Alibert}, {Udry}, \&
  {Mordasini}}]{mishra2023b}
---. 2023{\natexlab{b}}, \aap, 670, A69

\bibitem[{{Ostberg} {et~al.}(2023){Ostberg}, {Kane}, {Li}, {Schwieterman},
  {Hill}, {Bott}, {Dalba}, {Fetherolf}, {Head}, \& {Unterborn}}]{ostberg2023}
{Ostberg}, C., {Kane}, S.~R., {Li}, Z., {et~al.} 2023, \aj, 165, 168

\bibitem[{{Petigura} {et~al.}(2017{\natexlab{a}}){Petigura}, {Sinukoff},
  {Lopez}, {Crossfield}, {Howard}, {Brewer}, {Fulton}, {Isaacson}, {Ciardi},
  {Howell}, {Everett}, {Horch}, {Hirsch}, {Weiss}, \&
  {Schlieder}}]{petigura2017}
{Petigura}, E.~A., {Sinukoff}, E., {Lopez}, E.~D., {et~al.} 2017{\natexlab{a}},
  \aj, 153, 142

\bibitem[{{Petigura} {et~al.}(2017{\natexlab{b}}){Petigura}, {Howard}, {Marcy},
  {Johnson}, {Isaacson}, {Cargile}, {Hebb}, {Fulton}, {Weiss}, {Morton},
  {Winn}, {Rogers}, {Sinukoff}, {Hirsch}, \& {Crossfield}}]{CKS1}
{Petigura}, E.~A., {Howard}, A.~W., {Marcy}, G.~W., {et~al.}
  2017{\natexlab{b}}, \aj, 154, 107

\bibitem[{{Radovan} {et~al.}(2014){Radovan}, {Lanclos}, {Holden}, {Kibrick},
  {Allen}, {Deich}, {Rivera}, {Burt}, {Fulton}, {Butler}, \&
  {Vogt}}]{radovan2014}
{Radovan}, M.~V., {Lanclos}, K., {Holden}, B.~P., {et~al.} 2014, Society of
  Photo-Optical Instrumentation Engineers (SPIE) Conference Series, Vol. 9145,
  {The automated planet finder at Lick Observatory} (SPIE Press), 91452B

\bibitem[{{Rasio} \& {Ford}(1996)}]{rasio1996}
{Rasio}, F.~A., \& {Ford}, E.~B. 1996, Science, 274, 954

\bibitem[{{Rein} \& {Liu}(2012)}]{rein2012a}
{Rein}, H., \& {Liu}, S.~F. 2012, \aap, 537, A128

\bibitem[{{Rein} \& {Tamayo}(2015)}]{Rein2015}
{Rein}, H., \& {Tamayo}, D. 2015, \mnras, 452, 376

\bibitem[{{Ricker} {et~al.}(2015){Ricker}, {Winn}, {Vanderspek}, {Latham},
  {Bakos}, {Bean}, {Berta-Thompson}, {Brown}, {Buchhave}, {Butler}, {Butler},
  {Chaplin}, {Charbonneau}, {Christensen-Dalsgaard}, {Clampin}, {Deming},
  {Doty}, {De Lee}, {Dressing}, {Dunham}, {Endl}, {Fressin}, {Ge}, {Henning},
  {Holman}, {Howard}, {Ida}, {Jenkins}, {Jernigan}, {Johnson}, {Kaltenegger},
  {Kawai}, {Kjeldsen}, {Laughlin}, {Levine}, {Lin}, {Lissauer}, {MacQueen},
  {Marcy}, {McCullough}, {Morton}, {Narita}, {Paegert}, {Palle}, {Pepe},
  {Pepper}, {Quirrenbach}, {Rinehart}, {Sasselov}, {Sato}, {Seager},
  {Sozzetti}, {Stassun}, {Sullivan}, {Szentgyorgyi}, {Torres}, {Udry}, \&
  {Villasenor}}]{ricker2015}
{Ricker}, G.~R., {Winn}, J.~N., {Vanderspek}, R., {et~al.} 2015, Journal of
  Astronomical Telescopes, Instruments, and Systems, 1, 014003

\bibitem[{Rubenzahl {et~al.}(2021)Rubenzahl, Dai, Howard, Chontos, Giacalone,
  Lubin, Rosenthal, Isaacson, Batalha, Crossfield, Dressing, Fulton, Huber,
  Kane, Petigura, Robertson, Roy, Weiss, Beard, Hill, Mayo, Mocnik, Murphy, \&
  Scarsdale}]{Rubenzahl2021}
Rubenzahl, R.~A., Dai, F., Howard, A.~W., {et~al.} 2021, The Astronomical
  Journal, 161, 119

\bibitem[{Scarsdale {et~al.}(2021)Scarsdale, Murphy, Batalha, Crossfield,
  Dressing, Fulton, Howard, Huber, Isaacson, Kane, Petigura, Robertson, Roy,
  Weiss, Beard, Behmard, Chontos, Christiansen, Ciardi, Claytor, Collins,
  Collins, Dai, Dalba, Dragomir, Fetherolf, Fukui, Giacalone, Gonzales, Hill,
  Hirsch, Jensen, Kosiarek, de~Leon, Lubin, Lund, Luque, Mayo, Mo{\v{c}}nik,
  Mori, Narita, Nowak, Pall{\'{e}}, Rabus, Rosenthal, Rubenzahl, Schlieder,
  Shporer, Stassun, Twicken, Wang, Yahalomi, Jenkins, Latham, Ricker, Seager,
  Vanderspek, \& Winn}]{Scarsdale2021}
Scarsdale, N., Murphy, J. M.~A., Batalha, N.~M., {et~al.} 2021, The
  Astronomical Journal, 162, 215

\bibitem[{{Schlafly} \& {Finkbeiner}(2011)}]{schlafly2011}
{Schlafly}, E.~F., \& {Finkbeiner}, D.~P. 2011, \apj, 737, 103

\bibitem[{{Scott}(2019)}]{Scott2019}
{Scott}, N.~J. 2019, in AAS/Division for Extreme Solar Systems Abstracts,
  Vol.~51, AAS/Division for Extreme Solar Systems Abstracts, 330.15

\bibitem[{{Scott} {et~al.}(2018){Scott}, {Howell}, {Horch}, \&
  {Everett}}]{Scott2018}
{Scott}, N.~J., {Howell}, S.~B., {Horch}, E.~P., \& {Everett}, M.~E. 2018,
  \pasp, 130, 054502

\bibitem[{{Seager}(2010)}]{Seager2010}
{Seager}, S. 2010, {Exoplanets}

\bibitem[{{Smith} {et~al.}(2012){Smith}, {Stumpe}, {Van Cleve}, {Jenkins},
  {Barclay}, {Fanelli}, {Girouard}, {Kolodziejczak}, {McCauliff}, {Morris}, \&
  {Twicken}}]{smith2012d}
{Smith}, J.~C., {Stumpe}, M.~C., {Van Cleve}, J.~E., {et~al.} 2012, \pasp, 124,
  1000

\bibitem[{{STScI}(2018)}]{MAST}
{STScI}. 2018, TESS Input Catalog and Candidate Target List,  STScI/MAST,
  doi:10.17909/FWDT-2X66

\bibitem[{{Stumpe} {et~al.}(2012){Stumpe}, {Smith}, {Van Cleve}, {Twicken},
  {Barclay}, {Fanelli}, {Girouard}, {Jenkins}, {Kolodziejczak}, {McCauliff}, \&
  {Morris}}]{stumpe2012}
{Stumpe}, M.~C., {Smith}, J.~C., {Van Cleve}, J.~E., {et~al.} 2012, \pasp, 124,
  985

\bibitem[{Suzuki {et~al.}(2018)Suzuki, Bennett, Ida, Mordasini, Bhattacharya,
  Bond, Donachie, Fukui, Hirao, Koshimoto, Miyazaki, Nagakane, Ranc,
  Rattenbury, Sumi, Alibert, \& Lin}]{Suzuki_2018}
Suzuki, D., Bennett, D.~P., Ida, S., {et~al.} 2018, The Astrophysical Journal,
  869, L34

\bibitem[{{Trilling} {et~al.}(1998){Trilling}, {Benz}, {Guillot}, {Lunine},
  {Hubbard}, \& {Burrows}}]{trilling1998}
{Trilling}, D.~E., {Benz}, W., {Guillot}, T., {et~al.} 1998, \apj, 500, 428

\bibitem[{{Turtelboom} {et~al.}(2022){Turtelboom}, {Weiss}, {Dressing},
  {Nowak}, {Pall{\'e}}, {Beard}, {Blunt}, {Brinkman}, {Chontos}, {Claytor},
  {Dai}, {Dalba}, {Giacalone}, {Gonzales}, {Harada}, {Hill}, {Holcomb},
  {Korth}, {Lubin}, {Masseron}, {MacDougall}, {Mayo}, {Mo{\v{c}}nik}, {Akana
  Murphy}, {Polanski}, {Rice}, {Rubenzahl}, {Scarsdale}, {Stassun}, {Tyler},
  {Zandt}, {Crossfield}, {Deeg}, {Fulton}, {Gandolfi}, {Howard}, {Huber},
  {Isaacson}, {Kane}, {Lam}, {Luque}, {Mart{\'\i}n}, {Morello}, {Orell-Miquel},
  {Petigura}, {Robertson}, {Roy}, {Van Eylen}, {Baker}, {Belinski}, {Bieryla},
  {Ciardi}, {Collins}, {Cutting}, {Della-Rose}, {Ellingsen}, {Furlan}, {Gan},
  {Gnilka}, {Guerra}, {Howell}, {Jimenez}, {Latham}, {Larivi{\`e}re}, {Lester},
  {Lillo-Box}, {Luker}, {Mann}, {Plavchan}, {Safonov}, {Skinner}, {Strakhov},
  {Wittrock}, {Caldwell}, {Essack}, {Jenkins}, {Quintana}, {Ricker},
  {Vanderspek}, {Seager}, \& {Winn}}]{Turtelboom2022}
{Turtelboom}, E.~V., {Weiss}, L.~M., {Dressing}, C.~D., {et~al.} 2022, \aj,
  163, 293

\bibitem[{{Valenti} {et~al.}(1995){Valenti}, {Marcy}, \&
  {Basri}}]{valenti1995a}
{Valenti}, J.~A., {Marcy}, G.~W., \& {Basri}, G. 1995, \apj, 439, 939

\bibitem[{{Vogt} {et~al.}(1994){Vogt}, {Allen}, {Bigelow}, {Bresee}, {Brown},
  {Cantrall}, {Conrad}, {Couture}, {Delaney}, {Epps}, {Hilyard}, {Hilyard},
  {Horn}, {Jern}, {Kanto}, {Keane}, {Kibrick}, {Lewis}, {Osborne},
  {Pardeilhan}, {Pfister}, {Ricketts}, {Robinson}, {Stover}, {Tucker}, {Ward},
  \& {Wei}}]{vogt1994}
{Vogt}, S.~S., {Allen}, S.~L., {Bigelow}, B.~C., {et~al.} 1994, Society of
  Photo-Optical Instrumentation Engineers (SPIE) Conference Series, Vol. 2198,
  {HIRES: the high-resolution echelle spectrometer on the Keck 10-m Telescope}
  (SPIE Press), 362

\bibitem[{{Vogt} {et~al.}(2014){Vogt}, {Radovan}, {Kibrick}, {Butler},
  {Alcott}, {Allen}, {Arriagada}, {Bolte}, {Burt}, {Cabak}, {Chloros},
  {Cowley}, {Deich}, {Dupraw}, {Earthman}, {Epps}, {Faber}, {Fischer}, {Gates},
  {Hilyard}, {Holden}, {Johnston}, {Keiser}, {Kanto}, {Katsuki}, {Laiterman},
  {Lanclos}, {Laughlin}, {Lewis}, {Lockwood}, {Lynam}, {Marcy}, {McLean},
  {Miller}, {Misch}, {Peck}, {Pfister}, {Phillips}, {Rivera}, {Sand ford},
  {Saylor}, {Stover}, {Thompson}, {Walp}, {Ward}, {Wareham}, {Wei}, \&
  {Wright}}]{vogt2014a}
{Vogt}, S.~S., {Radovan}, M., {Kibrick}, R., {et~al.} 2014, \pasp, 126, 359

\bibitem[{{Weiss} {et~al.}(2018){Weiss}, {Marcy}, {Petigura}, {Fulton},
  {Howard}, {Winn}, {Isaacson}, {Morton}, {Hirsch}, {Sinukoff}, {Cumming},
  {Hebb}, \& {Cargile}}]{weiss2018a}
{Weiss}, L.~M., {Marcy}, G.~W., {Petigura}, E.~A., {et~al.} 2018, \aj, 155, 48

\bibitem[{{Weiss} {et~al.}(2021){Weiss}, {Dai}, {Huber}, {Brewer}, {Collins},
  {Ciardi}, {Matthews}, {Ziegler}, {Howell}, {Batalha}, {Crossfield},
  {Dressing}, {Fulton}, {Howard}, {Isaacson}, {Kane}, {Petigura}, {Robertson},
  {Roy}, {Rubenzahl}, {Twicken}, {Claytor}, {Stassun}, {MacDougall}, {Chontos},
  {Giacalone}, {Dalba}, {Mocnik}, {Hill}, {Beard}, {Akana Murphy}, {Rosenthal},
  {Behmard}, {Van Zandt}, {Lubin}, {Kosiarek}, {Lund}, {Christiansen},
  {Matson}, {Beichman}, {Schlieder}, {Gonzales}, {Brice{\~n}o}, {Law}, {Mann},
  {Collins}, {Evans}, {Fukui}, {Jensen}, {Murgas}, {Narita}, {Palle},
  {Parviainen}, {Schwarz}, {Tan}, {Acton}, {Bryant}, {Chaushev}, {Gill},
  {Eigm{\"u}ller}, {Jenkins}, {Ricker}, {Seager}, \& {Winn}}]{Weiss2021}
{Weiss}, L.~M., {Dai}, F., {Huber}, D., {et~al.} 2021, \aj, 161, 56

\bibitem[{{Wenger} {et~al.}(2000){Wenger}, {Ochsenbein}, {Egret}, {Dubois},
  {Bonnarel}, {Borde}, {Genova}, {Jasniewicz}, {Lalo{\"e}}, {Lesteven}, \&
  {Monier}}]{Simbad}
{Wenger}, M., {Ochsenbein}, F., {Egret}, D., {et~al.} 2000, \aaps, 143, 9

\bibitem[{{Winn} \& {Fabrycky}(2015)}]{winn2015}
{Winn}, J.~N., \& {Fabrycky}, D.~C. 2015, \araa, 53, 409

\bibitem[{{Zink} {et~al.}(2023){Zink}, {Hardegree-Ullman}, {Christiansen},
  {Petigura}, {Boley}, {Bhure}, {Rice}, {Yee}, {Isaacson}, {Fernandes},
  {Howard}, {Blunt}, {Lubin}, {Chontos}, {Pidhorodetska}, \&
  {MacDougall}}]{Zink2023}
{Zink}, J.~K., {Hardegree-Ullman}, K.~K., {Christiansen}, J.~L., {et~al.} 2023,
  \aj, 165, 262

\end{thebibliography}
\end{document}